\documentclass[prd,twocolumn,preprintnumbers,amsmath,amssymb,superscriptaddress,nofootinbib]{revtex4}
\pdfoutput=1

\usepackage{epsfig,psfrag,graphics,verbatim}
\usepackage{graphicx}% Include figure files
\usepackage{epstopdf}
\usepackage{dcolumn}% Align table columns on decimal point
\usepackage{bm}% bold math
\usepackage{slashed}
\usepackage{color}
\usepackage{amssymb}

\newcommand{\GeV}{\,\mbox{GeV}}

\begin{document} 
\preprint{DESY 12-120}
\preprint{TUM-HEP 847/12}

\title{Closing in on mass-degenerate dark matter scenarios\\ with antiprotons and direct detection}

\author{Mathias Garny}
\affiliation{Deutsches Elektronen-Synchrotron DESY, Notkestra\ss{}e 85, 22603 Hamburg, Germany}

\author{Alejandro Ibarra}
\affiliation{Physik-Department T30d, Technische Universit\"at M\"unchen, James-Franck-Stra\ss{}e, 85748 Garching, Germany}

\author{Miguel Pato}
\affiliation{Physik-Department T30d, Technische Universit\"at M\"unchen, James-Franck-Stra\ss{}e, 85748 Garching, Germany}

\author{Stefan Vogl}
\affiliation{Physik-Department T30d, Technische Universit\"at M\"unchen, James-Franck-Stra\ss{}e, 85748 Garching, Germany}

\date{\today}% It is always \today, today,

\begin{abstract}
Over the last years both cosmic-ray antiproton measurements and direct dark matter searches have proved particularly effective in constraining the nature of dark matter candidates. The present work focusses on these two types of constraints in a minimal framework which features a Majorana fermion as the dark matter particle and a scalar that mediates the coupling to quarks. Considering a wide range of coupling schemes, we derive antiproton and direct detection constraints using the latest data and paying close attention to astrophysical and nuclear uncertainties. Both signals are strongly enhanced in the presence of degenerate dark matter and scalar masses, but we show that the effect is especially dramatic in direct detection. Accordingly, the latest direct detection limits take the lead over antiprotons. We find that antiproton and direct detection data set stringent lower limits on the mass splitting, reaching 19\% at a 300 GeV dark matter mass for a unity coupling. Interestingly, these limits are orthogonal to ongoing collider searches at the Large Hadron Collider, making it feasible to close in on degenerate dark matter scenarios within the next years.
\end{abstract}

%\pacs{}% PACS, the Physics and Astronomy
%                             % Classification Scheme.
\maketitle

\section{Introduction}
\par Complementarity between different signals has always been highlighted as the preferred way forward in the search for weakly interacting massive particles (WIMPs) -- see \cite{BertoneReview,BergstromReview,BertoneBook,BergstromMulti} for general-purpose reviews. However, WIMP searches were for a long time an extremely data-starved field of research with experiments lagging far behind theoretical predictions. That is no longer the case. With large amounts of data pouring in from direct, indirect and collider dark matter (DM) searches, complementarity studies are now possible at an unprecedented level of detail \cite{Zheng:2010js,Bertone:2011nj,Bertone:2011pq,Strege:2011pk,Yu:2011by}. Direct searches, in particular, have developed tremendously over the last decade benefiting from the use of various targets and techniques to measure WIMP-induced nuclear recoils. Presently, the situation is rather blurry: while experiments such as DAMA/LIBRA \cite{DAMA2008,DAMA2008b,DAMA2010}, CoGeNT \cite{cogent,cogentannualmod} and CRESST \cite{CRESST2011} report excess of nuclear recoil events and also evidence for annual modulation in the case of DAMA/LIBRA and CoGeNT, other collaborations -- including XENON10/100 \cite{Xenon10,XENONSD,Xenon100,Xenon1002,Xenon100_2012}, CDMS \cite{CDMS2009,cdms10,Ahmed:2012vq}, SIMPLE \cite{SIMPLE10a,SIMPLE10b,SIMPLE11} or COUPP \cite{COUPP11,COUPP12} -- find null results. Leaving aside this controversy, that has been discussed at length in the literature \cite{ChangIso,HooperCoGeNT,FengIso,Arina2011,Schwetz:2011xm,Kopp:2011yr}, it is clear that the whole array of different targets used boasts huge sensitivities to both spin-dependent (SD) and spin-independent (SI) WIMP-nucleus scattering. Together with direct detection, indirect searches via antiprotons have provided valuable hints in shaping the phenomenology of viable WIMP models. In fact, the exquisite data on the cosmic-ray antiproton flux collected by experiments such as BESS \cite{Orito:1999re,Abe:2008sh,BESSPolarII} or PAMELA \cite{Adriani:2010rc} fall nicely on top of the expectations from cosmic-ray spallations in the Galaxy. This means in practice that dark matter annihilations or decays cannot yield copious fluxes of antiprotons, and correspondingly the coupling to quarks is much constrained \cite{Donato:2008jk,Kappl:2011jw}. Notice that this is precisely the coupling that drives WIMP-nucleus scattering in underground detectors, which makes antiprotons and direct searches particularly suitable for complementarity studies.

\par Several complications arise when exploring the complementarity between two or more observables. Firstly, if we are to draw sound conclusions, all relevant uncertainties must be treated carefully. In the case at hand, the unknowns are sizable and different in nature: on the one hand, antiproton searches are prone to uncertainties on the galactic dark matter profile \cite{Navarro:2003ew,Gao:2007gh,Diemand:2008in,Navarro:2008kc} and on cosmic-ray propagation \cite{SM98,Maurin:2001sj,Delahaye07,dragon2,Putze1,Trotta:2010mx,Evoli:2011id}; on the other hand, direct searches suffer from the lack of knowledge on the local dark matter density \cite{CaldwellOstriker,Gates1995,UllioBuckley,BelliFornengo,CatenaUllio,Weber:2009pt,SaluccilocalDM,paperDMlocal,papermuLens,Bovy:2012tw} and local velocity distribution \cite{Smith:2006ym,Xue2008,Sofue2008,Reid2009,Bovy2009,McMillan2009,Lisanti:2010qx} as well as from nuclear uncertainties \cite{Pumplin:2002vw,EllisDD}. Secondly, combining signals such as antiprotons and direct detection requires the specification of an underlying particle physics framework. For instance, the author of references \cite{Lavalle:2010yw,Lavalle:2011fm} points out the usefulness of antiproton measurements to constrain light WIMPs able to accommodate the hints of signal in DAMA/LIBRA and CoGeNT. This claim is, though, dependent on how dark matter couples to quarks, especially the light quarks abundant in nucleons that get struck in direct detection experiments. A definitive conclusion is therefore necessarily model-dependent. This highlights the difficulty in pursuing complementarity studies and deriving conclusive results.

\par Now, model building imposes hardly any boundaries on the phenomenology of dark matter candidates. Nevertheless, one can learn a lot by focussing on relatively simple realisations. For example, it has been recently emphasized \cite{Hisano:2011um,Garny:2011cj,Garny:2011ii,Asano:2011ik} that the presence of mediator particles degenerate in mass with the WIMP leads to enhanced signals in both antiprotons and direct detection. In the case of the former, degeneracy boosts the contribution of $2\to3$ processes to the annihilation yields \cite{Bergstrom:1989jr,Flores:1989ru,Drees:1993bh,Garny:2011cj,Garny:2011ii,Asano:2011ik}, while in the latter it induces almost-resonant scattering in SD and SI interactions \cite{Hisano:2011um}. Besides, all this proceeds at mass degeneracies below the trigger of searches at the Large Hadron Collider (LHC). To the best of our knowledge, the complementarity between antiprotons and direct searches has not been conveniently explored in this important case of degenerate mass states. The present work is precisely devoted to fill that gap, and is a first step in exploring realistically the interplay between direct and indirect searches within mass-degenerate WIMP frameworks. In particular, we show explicitly how data already at hand effectively constrain the presence of degenerate mass states, and how a reasonable experimental effort over the coming years will be able to close in on these scenarios. As we shall see, collider searches are orthogonal to antiproton and direct detection, and hence a multi-disciplinary approach is highly desirable for the near future of dark matter searches. 

\par We start by introducing the main ingredients of our particle physics framework in Section \ref{sec:model}. Then, Sections \ref{sec:Indirect} and \ref{sec:DD} are devoted to the formalism behind antiproton and direct detection constraints, respectively. Taking special care in modelling all relevant uncertainties, we derive in Section \ref{sec:res} the limits imposed by the latest SD and SI direct detection searches as well as cosmic-ray antiproton data, before concluding in Section \ref{sec:conc}.

\section{Particle physics model}\label{sec:model}

\par Direct detection experiments probe mainly the interaction of dark matter with the light quarks, the fundamental scattering process being $\chi q\to \chi q$. As is well-known, the corresponding annihilation process $\chi\chi\to q\bar q$ that is obtained from crossing the scattering diagram is strongly suppressed for Majorana fermions, either by the quark mass or the velocity $v\sim 10^{-3}c$ in the Galactic halo. Therefore, at first sight, it may seem hopeless to constrain the coupling of Majorana dark matter to light quarks via indirect detection. However, the helicity suppression can be lifted by the additional radiation of a spin-one boson in the final state~\cite{Bergstrom:1989jr,Flores:1989ru}. Because of this property, the two-to-three annihilation channels $\chi\chi\to q\bar q V$, with $V$ being a gluon~\cite{Flores:1989ru,Drees:1993bh,Barger:2006gw}, a $Z/W$-boson~\cite{Garny:2011ii,Garny:2011cj,Ciafaloni:2011sa,Ciafaloni:2011gv,Ciafaloni:2012gs,Bell:2011if,Bell:2011eu} (see also \cite{Kachelriess:2009zy,Ciafaloni:2010ti}) or a photon~\cite{Bergstrom:1989jr,Flores:1989ru,Bringmann:2007nk} typically dominate over the two-to-two process, provided that the particle mediating the annihilation has a mass that is of the same order as the dark matter mass~\cite{Garny:2011cj}. In particular, the emission of gluons leads to a rather efficient production of antiprotons, making it possible to obtain limits on the dark matter coupling to light quarks from indirect detection, and consequently to relate direct and indirect detection constraints in an unambiguous manner.

\par In the following, we consider a simple model for mass-degenerate dark matter that is suitable for a comparative analysis of direct and indirect detection constraints. It consists of a minimal extension of the Standard Model (SM) by a Majorana fermion $\chi$, which constitutes the dark matter in the Universe, and a scalar particle $\eta$ not much heavier than $\chi$, which provides the portal to the Standard Model via a Yukawa interaction with the light quarks~\cite{Garny:2011ii}. Although this simplified model is not uniquely related to a particular framework of physics beyond the SM, its particle content is motivated by the minimal supersymmetric extension of the Standard Model (MSSM).

\subsection{Outline of the model}

\par The dark matter particle $\chi$ is taken to be a singlet under the Standard Model gauge group, while $\eta$ is a singlet under $SU(2)$, a triplet under $SU(3)$ and carries hypercharge such that the electric charge of the quarks is obtained. The total Lagrangian of the model can be written as:
\begin{multline}
\mathcal{L}= \mathcal{L}_{SM} + {\textstyle\frac12} \bar \chi^c i\slashed {\partial} \chi
    -{\textstyle\frac12}m_\chi \bar \chi^c\chi + (D_\mu
    \eta)^\dagger  (D^\mu \eta) \\
 -m_\eta^2 \eta^\dagger\eta + \mathcal{L}_{int}. 
\end{multline}
Here, $\mathcal{L}_{SM}$ denotes the Standard Model Lagrangian, and $D_{\mu}$ is the usual covariant derivative. The remaining part of the Lagrangian,  $\mathcal{L}_{int}$, contains the interactions of the new particles with the quarks.
In the following, we assume that $\eta$ carries a flavour quantum number, thus ensuring that $\eta$ couples only to one quark flavour and does not induce any flavour changing effects. 
To be more precise we consider the cases where $\eta$ has $U=-1$, $D=-1$, $S=-1$ or $B=-1$ as well as a model with three mass-degenerate scalars $\eta_{u,d,s}$, 
each of which couples to one of the light quarks. Thus the interaction Lagrangian can be expressed as
\begin{align}
  {\cal L}_{int} &= -f \bar \chi \Psi_R \eta+{\rm h.c.} \;,
\label{eq:singlet-eR}
\end{align}
where $\Psi= u,\, d, \, s$, or $b$. Similarly, one could also consider an analogous model where dark matter couples to the left-handed quarks. However, we do not discuss this case separately here since it yields very similar constraints.

\par Formally, the simplified model described above can be obtained within the MSSM in the limiting case of a bino as the lightest supersymmetric particle, and a right-handed squark as the next-to-lightest one. The Yukawa coupling $f=f_{SUSY}$ is then fixed in terms of the $U(1)_Y$ gauge coupling $g'$ to $f_{SUSY}=4g'/(3\sqrt{2})\sim 0.33$ for up-type quarks and $f_{SUSY}=2g'/(3\sqrt{2})\sim 0.16$ for down-type quarks (for the analogous model involving left-handed quarks, one would have $f_{SUSY}=g'/(3\sqrt{2})$). In order to explore the full available parameter space, we shall consider the coupling $f$ as a free parameter in the following.
The quantum numbers of the scalar particle allow an additional renormalizable coupling
to the Higgs field, ${\cal L}_{int}^h=\lambda_3 H^\dag H \eta^\dag\eta$. It provides an additional mass term for $\eta$,
which can be absorbed into the tree-level mass, $m_\eta^2+\lambda_3 v_{EW}^2\to m_\eta^2$. Apart from that, it gives rise to couplings of $ \eta $ to the Higgs boson, however since only the couplings of the dark matter particle $ \chi $ are relevant for direct and indirect detection this turns out to be irrelevant for these limits. The most fundamental constraint on light coloured particles comes from the non-observation of an excess in the invisible decay width of the Z boson at LEP, $ \Delta \Gamma_{inv} < 2.0 $ MeV \cite{TheALEPHCollaborationPhys.Rept.427:257-4542006}. This excludes exotic charged particles with a mass below $40$ GeV  \cite{Nakamura:2010zzi}, and since we are interested in the case of a compressed mass spectrum we only consider dark matter masses $m_{\chi} \gtrsim 40$ GeV  throughout this analysis.   

\subsection{Thermal freeze-out}

\par
\begin{figure*}[t]
 \begin{center}
  \includegraphics[width=0.9 \textwidth]{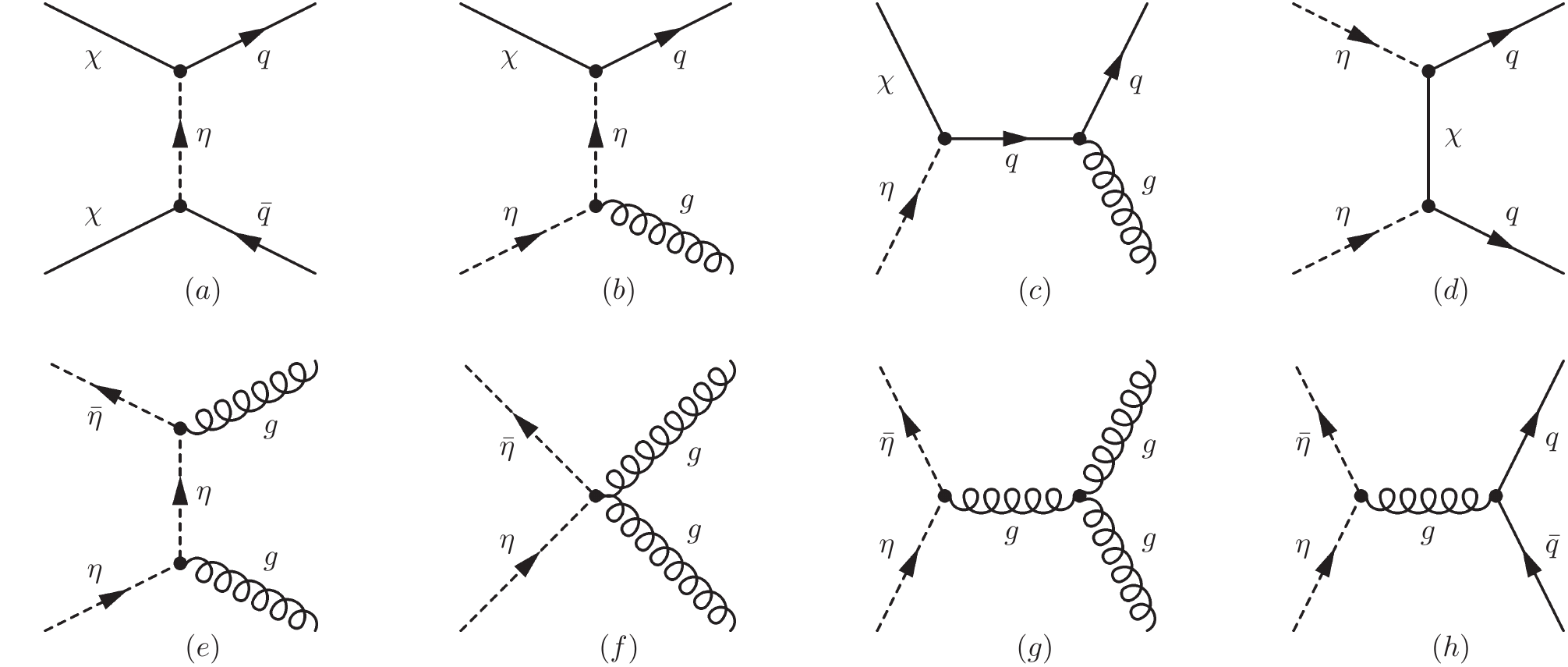}
 \end{center}
  \caption{Feynman diagrams showing the leading (co-)annihilation processes that contribute to the thermal freeze-out density. Two additional diagrams corresponding to $(a)$ with reversed charge flow and to $(d)$ with flipped legs are not shown, as well as several diagrams for charge conjugated processes.}
  \label{fig:coannihilation}
\end{figure*}

 One of the main arguments in favor of a WIMP as a dark matter candidate is that such a particle can be produced  quite naturally by thermal freeze-out in the early Universe. 
The model and the parameter space of interest here, i.e.~a compressed mass spectrum of the dark matter and the next-to-lightest particle beyond the Standard Model, make it necessary to treat thermal production 
with special care, since coannihilation can induce great deviations of the relic abundance from generic expectations  \cite{Griest:1990kh}.  In order to include all the relevant interactions of $\eta$ and $\chi$ we  choose to use a fully numerical calculation of thermal freeze-out by MicrOMEGAS \cite{Belanger:2010gh} which we have checked against a semi-analytic approximation~\cite{Ellis:2001nx}. The analytic result for the freeze-out density can be expressed as
\begin{multline}\label{eq:omegaDM}
\Omega_{DM} h^2\simeq 0.11\, \frac{\langle\sigma_0 v\rangle}{\langle\sigma_{eff} v\rangle}\,\frac{T_{f0}}{T_f} \, \left(\frac{0.35}{f}\right)^4 \frac{1}{N_c} \\
\times \left(\frac{m_{\chi}}{100\mbox{GeV}}\right)^2  \frac{(1+m_{\eta}^2/m_{\chi}^2)^4}{1+m_{\eta}^4/m_{\chi}^4} \;,
\end{multline}
where $N_c=3$ is a colour factor, and $\sigma_0 v$ as well as $T_{f0}\simeq m_\chi/25$ correspond to the annihilation cross-section and freeze-out temperature without taking coannihilations into account \cite{Garny:2011ii,Garny:2011cj}, respectively. The effective cross-section is given by
\begin{equation}
\sigma_{eff} v = \sum_{ij} \frac{n^{eq}_i n^{eq}_j}{(\sum_k n^{eq}_k)^2} \sigma_{ij} v\,,
\end{equation}
with $n^{eq}_i=g_i(m_iT/(2\pi))^{3/2}e^{-m_i/T}$ and $g_\chi=2$, $g_\eta=g_{\bar\eta}=1$. When expanding $\sigma v=a+bv^2$ into s- and p-wave contributions, the thermal average is given by $\langle\sigma v\rangle =a+3b T/m_\chi$. 
The Feynman diagrams for the leading processes which we take into account are shown in Fig.~\ref{fig:coannihilation}.
The cross-section for the annihilation of dark matter particles, $\chi\chi\to q\bar q$, is helicity- and/or velocity-suppressed,
\begin{equation}\label{eq:sigv2to2}
\sigma v(\chi\chi\to q\bar q)  =  \frac{3f^4m_q^2}{32\pi(m_\chi^2+m_\eta^2)^2} + \frac{f^4 v^2}{16\pi m_\chi^2} \frac{1+m_\eta^4/m_\chi^4}{(1+m_\eta^2/m_\chi^2)^4} \,,
\end{equation}
while the cross-sections for the processes $\chi\eta\to qg$, $\eta\eta\to qq$, and $\eta\bar\eta\to gg$ have a non-zero s-wave contribution for $m_q\to 0$,
\begin{eqnarray}
\sigma v(\chi\eta\to q g) & = & \frac{f^2g_s^2}{24\pi}\frac{1}{m_\eta(m_\eta+m_\chi)}\,, \nonumber\\
\sigma v(\eta\bar\eta\to gg) & = & \frac{7g_s^4}{216 \pi m_\eta^2}\,,\quad  \nonumber\\
\sigma v(\eta\eta\to qq) \ & = & \ \frac{ f^4}{6\pi}\frac{m_\chi^2}{(m_\chi^2+m_\eta^2)^2}\,.
\end{eqnarray}
The s-wave contribution for $\eta\bar\eta\to q\bar q$ (see Fig.~\ref{fig:coannihilation}, diagram ($h$) as well as ($d$) with flipped charge flow in one of the lines) is helicity-suppressed, and therefore less important. For any given mass spectrum, we fix the coupling $f$ such that the relic density matches the Wilkinson Microwave Anisotropy Probe (WMAP) value \cite{Komatsu2010}. For a large mass splitting, $m_\eta-m_\chi\gg T_{f0}$, coannihilations are suppressed, and the resulting coupling $f$ is of order $1/3$ for dark matter masses in the $100$ GeV range. For smaller mass splittings, additional coannihilation channels contribute, and therefore a smaller coupling $f$ is sufficient to achieve the same relic density. Below a certain value for the dark matter mass and the mass splitting, the contribution of the annihilation channel $\eta\bar\eta\to gg$, whose cross-section depends only on the strong coupling, can become so large that the freeze-out density falls below the WMAP value even for\footnote{We assume that inelastic scattering processes like $\chi q \leftrightarrow \eta g$ (as well as a similar crossed process) are fast enough to maintain chemical equilibrium among $\eta$ and $\chi$ during thermal freeze-out in the case that $m_\eta-m_\chi\lesssim T_{f0}$. This assumption is necessary for applying the conventional treatment of freeze-out with coannihilations~\cite{Griest:1990kh}. For our scenario, it is well-justified provided that $f \gtrsim 10^{-4}$, which we assume to be the case.} $f\ll g_s$. For a mass splitting $m_\eta/m_\chi=1.1~(1.01)$, this leads to a lower limit on the dark matter mass of $m_\chi\gtrsim \mathcal{O}[200~(1000)\GeV]$, for which thermal freeze-out can account for the observed dark matter density. This also explains the behaviour of the thermal relic density constraint shown in Fig.~\ref{fig:coupling} below. 
The results of the  numerical solution and the approximation are in good agreement, thus indicating that both approaches yield a reasonable description of thermal freeze-out. We have also checked that the inclusion of diagrams similar to the ones shown in Fig.~\ref{fig:coannihilation}, but with gluons replaced by electroweak gauge bosons or Higgs bosons, leads to minor corrections, assuming $|\lambda_3|\lesssim 1$ for the coupling to the Higgs. We also checked that annihilation into top quarks and into a pair of Higgs bosons has no sizeable impact. In the following we prefer to use the limits obtained by MicrOMEGAS since they are expected to be more precise.  

\section{Antiproton constraints}\label{sec:Indirect}

\par In this section we first discuss the annihilation processes that are relevant for indirect detection, and then review constraints arising from the primary flux of antiprotons induced by the decay of annihilation products. As discussed above, the lowest order annihilation process $\chi\chi\to q\bar q$, see Eq.~(\ref{eq:sigv2to2}), is strongly suppressed. In particular, since the typical dark matter velocity in the halo of the Milky Way is of the order $v\sim 10^{-3}c$, the p-wave contribution to the cross-section is suppressed much more strongly than during thermal freeze-out. On the other hand, the s-wave contribution is suppressed by the masses of the light quarks, and therefore even smaller. Hence, $2\to 3$ annihilation processes, where the quark pair is accompanied by a gauge boson, typically yield the dominating contribution to the annihilation within our Galaxy. The differential cross-section for these processes is given by~\cite{Garny:2011ii}
\begin{multline}
\frac{vd\sigma(\chi\chi\to  q\bar q V)}{dE_V dE_q}  =  \frac{C_{V} f^4}{8\pi^2 m_{\chi}^4 } \, \\
\times \frac{(x_{max}-x)[(1-x-y)^2+(1-y)^2-M_V^2/(2m_\chi^2)]}{(1-2y-m_{\eta}^2/m_{\chi}^2)^2(3-2x-2y+m_{\eta}^2/m_{\chi}^2)^2} \;,
\end{multline}
where $x=E_V/m_{\chi}$, $y=E_q/m_{\chi}$, and $x_{max}=1+M_V^2/(4m_{\chi}^2)$. The factor for the various channels reads $C_\gamma=3q_f^2\alpha_{em}$, $C_Z=3q_f^2\tan^2(\theta_W)\alpha_{em}$, and $C_g=4\alpha_s$, where $q_f=2/3$ for up-type and $-1/3$ for down-type quarks, $\alpha_{em}=e^2/(4\pi)$, and $\alpha_s=g_s^2/(4\pi)$.
Since the gauge boson has to carry away one unit of angular momentum in order to lift the helicity suppression, its spectrum is rather hard and features a pronounced peak close to the dark matter mass~\cite{Garny:2011cj,Bringmann:2011ye}. When $V$ is a photon, this process of internal bremsstrahlung can therefore lead to a characteristic feature in the gamma-ray spectrum that may be observable in current and future gamma-ray telescopes such as Fermi LAT or the planned Cherenkov Telescope Array (CTA)~\cite{Bringmann:2011ye,Bringmann:2012vr}. In the present context, we concentrate on the antiproton flux, that is mainly produced by the annihilation into gluons, $\chi\chi\to q\bar qg$.

\subsection{Formalism \& uncertainties}

\par The annihilation of dark matter particles in the halo of the Milky Way induces a primary flux of
antiprotons that is potentially observable at the Earth. The rate of antiprotons produced at
position $\vec r$ with respect to the center of our Galaxy per unit of kinetic energy and volume is given by
\begin{align}
Q(T,\vec r)=\frac{1}{2}\frac{\rho^2(\vec r)}{m^2_\chi}
\sum_j \langle \sigma v\rangle_j \frac{dN^j_{\bar p}}{dT}\;,
\label{eq:source}
\end{align}
where $\langle \sigma v\rangle_j$ is the thermally averaged cross-section multiplied by the velocity in the annihilation channel $j$, $\rho(\vec r)$ is the distribution of dark matter particles in the Milky Way, and $dN^j_{\bar p}/dT$ is the energy spectrum of antiprotons produced in  that channel per unit of kinetic energy. We assume for simplicity a spherically symmetric dark matter distribution, and calculate the antiproton flux assuming a radial dependence given by either the isothermal, Navarro-Frenk-White (NFW) \cite{Navarro:1995iw,Navarro:1996gj} or Einasto profiles \cite{Navarro:2003ew,Graham:2006ae}, with scale radii $r_s=4.38,~24.42$ and  $28.44\,{\rm kpc}$~\cite{Garny:2011cj,Garny:2011ii}, respectively, and normalized to a local dark matter density $\rho_0\equiv \rho(r_\odot)=0.4 \,{\rm GeV}/{\rm cm}^3$~\cite{CatenaUllio,Weber:2009pt,SaluccilocalDM,papermuLens}. The spectrum of antiprotons is obtained using the event generator CalcHEP \cite{Pukhov:1999gg, Pukhov:2004ca} interfaced with PYTHIA 8.1 \cite{Sjostrand:2007gs}.

\par The largest source of uncertainty arises from the propagation of antiprotons from the production point to the Earth. Following \cite{ACR}, we use a stationary two-zone diffusion model with cylindrical boundary conditions. The number density of antiprotons per unit kinetic energy, $f_{\bar p}(T,\vec{r},t)$, approximately satisfies the following transport equation in this model:
\begin{multline}
0=\frac{\partial f_{\bar p}}{\partial t}=
\nabla \cdot (K(T,\vec{r})\nabla f_{\bar p})
-\nabla \cdot (\vec{V_c}(\vec{r})  f_{\bar p})\\
-2 h \delta(z) \Gamma_{\rm ann} f_{\bar p}+Q(T,\vec{r})\;.
\label{transport-antip}
\end{multline}
The boundary conditions require the solution $f_{\bar p}(T,\vec{r},t)$ to vanish at the boundary of the diffusion zone, which is approximated by a cylinder with half-height $L = 1-15~\rm{kpc}$ and radius $ R = 20 ~\rm{kpc}$. The influence of the galactic magnetic field is modeled by an energy-dependent diffusion term, $K(T)=K_0\beta{\cal R}^\delta$, where $\beta=v/c$ is related to the velocity and ${\cal R}=p(\textrm{GeV})/|Z|$ is the rigidity given by the ratio of momentum and electric charge.
In addition, the second term accounts for the drift induced by the Galactic wind, described by  $\vec{V}_c(\vec{r})=V_c\; {\rm sign}(z)\; \vec{k}$. Here $\vec{k}$ denotes the unit vector in the $z$-direction perpendicular to the Galactic plane.  The third term accounts for the  annihilation of antiprotons when interacting with ordinary matter in the Galactic disk. The annihilation rate is taken from~\cite{Garny:2011cj}. The flux arriving at the solar system is related to the number density by $\Phi^{\rm{IS}}_{\bar p}(T) = \frac{v}{4 \pi} f_{\bar p}(T,r_\odot)$. Finally, the flux observed at the top of the Earth atmosphere is influenced by the effect of solar modulation, which we model within the force field approximation~\cite{solarmodulation1,solarmodulation2,perko}, as detailed in~\cite{Garny:2011cj}. Since the PAMELA data used in this analysis was taken near solar minimum activity, we choose $\phi_F=500$\,MV for the solar modulation parameter for our numerical analysis. In order to take the uncertainties related to propagation into account, we use three sets of parameters, compatible with the cosmic boron-to-carbon flux ratio~\cite{Maurin:2001sj,Bringmann:2006im}, corresponding to minimum, medium and maximum antiproton flux with parameters as given e.g.~in \cite{Garny:2011cj}. 

\subsection{Experimental data \& background}

\par The most recent measurements of the antiproton flux by the PAMELA satellite experiment~\cite{Adriani:2010rc} in the energy range from $60$\,MeV to $180$\,GeV are in accordance with measurements performed by BESS~\cite{Orito:1999re,Abe:2008sh} ($0.1-4.2$\,GeV), IMAX~\cite{Mitchell:1996bi} ($0.25-3.2$\,GeV), WiZard/CAPRICE~\cite{Boezio:1997ec} ($0.62-3.19$\,GeV), and AMS \cite{Aguilar:2002ad} ($0.2-4$\,GeV). In the future, data from the AMS-02 experiment are expected to yield further information on the cosmic antiproton flux~\cite{Malinin:2007zz,Pato:2010ih}. The measured flux as well as the antiproton-to-proton ratio agree well with the expectations from secondary production of antiprotons from spallation of cosmic-ray nuclei, mainly protons and helium, on the interstellar medium \cite{Bringmann:2006im,Evoli:2011id}. This allows to set stringent upper limits on a possible primary contribution generated from dark matter annihilations. In our analysis we use the background flux calculated in Ref.~\cite{Donato:2001ms} based on the two-zone diffusion model, taking into account p-p, He-p, p-He and He-He nuclear reactions. The main uncertainties arise from the diffusion parameters and the nuclear cross-sections, and are estimated to be in the range of $10-25$\% depending on the energy. In contrast, the uncertainty stemming from the knowledge of the flux of cosmic nuclei and the composition of the interstellar medium are found to be subdominant. In order to obtain constraints on the dark matter annihilation cross-section, we compute the antiproton-to-proton ratio $ \bar p/p \equiv (\Phi_{\bar p}^{\rm sig} +  \Phi_{\bar p}^{\rm bkg} )/\Phi_p $ using the proton flux of \cite{Bringmann:2006im}. In order to obtain a conservative exclusion bound we adopt the minimal value for the antiproton background as discussed in \cite{Donato:2001ms} and a lower cut $T>1.5$ GeV on the kinetic energy. Upper limits on the Yukawa coupling  $f$, that controls the size of the annihilation cross-sections, are then obtained from the PAMELA $\bar p/p$ data \cite{Adriani:2010rc} using a $\chi^2$-test at $95\%$ confidence level (CL).

\section{Direct detection}\label{sec:DD}

\subsection{Formalism \& uncertainties}
\par The subject of direct detection of dark matter has been extensively covered in the literature \cite{LewinSmith,Jungman,Munoz:2003gx,BertoneBook}, so here we limit ourselves to outline the features of interest for our study following closely \cite{Pato:2010zk,Pato:2011de}. Given a target nucleus $N(A,Z)$, the differential rate of nuclear recoils induced by WIMPs is readily obtained by convoluting the local dark matter flux and the WIMP-nucleus scattering cross-section. We take the local WIMP velocity distribution to be a truncated Maxwellian with escape velocity $v_{esc}$ and circular velocity $v_0$. The SI form factor is implemented as in \cite{LewinSmith}, while the SD one is modelled following \cite{BednyakovFormFactor} whenever possible or using the simplified prescription of \cite{LewinSmith} otherwise. The nuclear spin expectation values can be found in \cite{BednyakovSpin} for various nuclei (for nuclei absent in that reference we use \cite{Giuliani:2005bd}). Finally, the remaining ingredients to specify in order to obtain the recoil rate are the effective SD and SI WIMP-nucleon couplings, $a_{p,n}$ and $f_{p,n}$.

\par At the microscopic level, spin-dependent WIMP-nucleus scattering is induced by the exchange of a scalar in the $s$-channel or a $Z$ boson in the $t$-channel between the WIMP and a quark \cite{Jungman}. In the framework of the class of models introduced in Section \ref{sec:model}, the only scalar at play is $\eta$ and there is no WIMP coupling to the $Z$ boson. Accordingly, the effective axial-vector WIMP-quark coupling reads
\begin{equation}\label{dq}
d_q = \frac{1}{8}\frac{f^2}{m_\eta^2-\left(m_\chi+m_q\right)^2} \quad ,
\end{equation}
and the effective spin-dependent WIMP-proton coupling is
\begin{equation}\label{ap}
a_p = \sum_{q=u,d,s}{ \frac{d_q}{\sqrt{2} G_F} \Delta q^{(p)}  }
\end{equation}
and likewise for $a_n$, where the parameters $\Delta q^{(p,n)}$ set the spin content of the nucleon.

\par Spin-independent scattering, on the other hand, results from WIMP-quark and WIMP-gluon interactions via the exchange of scalars and/or Higgses at both tree and loop level \cite{Jungman}. Specialising to the particle physics realisation in study -- which features the scalar $\eta$ only and no significant WIMP-Higgs coupling --, the effective spin-independent WIMP-proton coupling is
\begin{multline}\label{fp}
\frac{f_p}{m_p} = - \frac{m_\chi}{2} \sum_{q=u,d,s}{ f_{T_q}^{(p)}g_q} - \frac{8\pi}{9} b f_{TG}^{(p)} \\
- \frac{3}{2}m_\chi \sum_{q=u,d,s,b}{ g_q \left( q^{(p)}(2) + \bar{q}^{(p)}(2) \right)  } \quad ,
\end{multline}
where $f_{T_q}^{(p)}$ and $f_{TG}^{(p)}$ parametrise the quark and gluon content of the proton respectively, and $q^{(p)}(2),\bar{q}^{(p)}(2)$ are the second moments of the parton distribution functions. A similar expression holds for $f_n$. The effective WIMP couplings $g_q$ and $b$ are given by
\begin{eqnarray}\nonumber
g_q &=& - \frac{1}{8}\frac{f^2}{\left(m_\eta^2-\left(m_\chi+m_q\right)^2\right)^2} \quad , \\ \label{fqgqb}
b &=& \left(B_S -\frac{m_\chi}{2} B_{2S} - \frac{m_\chi^2}{4}B_{1S}\right) \propto f^2 \quad ,
\end{eqnarray}
in which $B_{\times }$ are loop amplitudes that can be found in \cite{DreesNojiri,Jungman}. In Eq.~\eqref{fp}, the first term corresponds to WIMP-quark scattering through the exchange of $\eta$ in the $s$-channel, the second term encloses the contribution of WIMP-gluon scattering at one-loop level involving quarks and $\eta$, and the last term includes the twist-2 operator part. The eventual contributions arising from couplings to the charm- and top-quarks are left out since they are absent of our scheme of couplings described in Section \ref{sec:model}.

\par It is clear from the framework outlined above that the presence of a mass state $\eta$ almost degenerate with the dark matter particle strongly enhances both SD and SI direct detection rates. Indeed, the SD effective coupling $d_q$ and the SI term $g_q$ in Eqs.~\eqref{dq} and \eqref{fqgqb} exhibit a resonance at $m_\eta=m_\chi+m_q$. This was the motivation for the approach followed in \cite{Hisano:2011um} and is further pursued in the present work. In order to avoid modelling the width of the exchanged scalar $\eta$ -- and thus introducing more freedom in our particle physics framework --, we restrict ourselves to mass splittings not too close to the resonance, namely $(m_\eta-m_\chi)/m_q\geq2$ for each quark coupling considered. Also, we shall only derive constraints for $m_\eta-m_\chi> 1\textrm{ GeV}\sim m_p$ where the expansion leading to the effective Lagrangian with couplings given by Eqs.~\eqref{dq} and \eqref{fqgqb} is valid. Even with such conservative restrictions, the amount of enhancement produced in direct detection by a compressed mass spectrum is very significant as shown in Section \ref{sec:res}.

\par Before proceeding with setting constraints, it is instructive to discuss the direct detection phenomenology of the dark matter models under consideration. First of all, we comment on the relative strength of each of the three SI terms in Eq.~\eqref{fp}. Generically speaking, for the scheme of possible coupling configurations discussed in Section \ref{sec:model} the $b$-term is always subdominant except for bottom-quark couplings. For the up-quark, down-quark and democratic light quarks couplings, the last term in Eq.~\eqref{fp} -- i.e.~the twist-2 contribution -- plays the major role, while for strange-quark couplings it is the first term due to scalar exchange that dominates. Despite this sort of interplay, the scalar exchange and twist-2 contributions are roughly of the same order of magnitude, which makes direct detection constraints almost independent of the coupling scheme (for the case of light quarks). Secondly, the ratio of SD to SI WIMP-proton cross-sections $\sigma_p^{SD}/\sigma_p^{SI}$ is easily much larger than unity, reaching $\sim 10^2-10^8$ for the up-quark coupling with $m_\eta/m_\chi=1.1$ and using mean values for the nuclear parameters (see below). Although the SD WIMP-nucleon cross-section is easily larger than its SI counterpart, it must be recalled that in direct detection SD sensitivities lag far behind SI searches. Actually, the strongest direct detection constraints on our DM model will be largely coming from SI couplings. Finally, let us point out that fiducial values are $a_n/a_p=-0.5$ and $f_n/f_p=0.6$ in the case of up-quark couplings, and $f_n/f_p=1$ in the case of bottom-quark couplings. These values depend on the adopted nuclear parameters, but are largely independent of $m_\eta$ and $m_\chi$.

\par Ultimately, our aim is to link the nuclear recoil event rate observed in a given experiment to the couplings and masses of the DM model. To pursue this path close attention must be paid to two kinds of inputs: astrophysical quantities and nuclear parameters. On the astrophysical side, the key quantities are the local dark matter density $\rho_0$ and the local velocity distribution which is regulated by the circular and escape velocities, $v_0$ and $v_{esc}$. We use the following fiducial values for these quantities \cite{Pato:2010zk}:
\begin{eqnarray}\label{astro}
\rho_0&=&0.4\textrm{ GeV/cm}^3 \, , \nonumber \\
v_0&=&230\pm 30 \textrm{ km/s} \, , \nonumber \\
v_{esc}&=&544 \textrm{ km/s} \quad .
\end{eqnarray}
The largest astrophysical uncertainties under our framework regard $\rho_0$ and $v_0$. Since the former affects similarly direct detection and antiproton constraints, we opt to gauge its uncertainty out and stick to the typical value 0.4 GeV/cm$^3$ \cite{CatenaUllio,Weber:2009pt,SaluccilocalDM,papermuLens}. The uncertainty on $v_0$ reported above is a reasonable assessment of the present (lack of) knowledge on this important galactic parameter \cite{Pato:2010zk,Xue2008,Sofue2008,Reid2009,Bovy2009,McMillan2009} and shall be used throughout to represent the astrophysical uncertainty in direct detection. Other astrophysical parameters as $v_{esc}$ are also uncertain to a certain level \cite{Smith:2006ym}, but such uncertainty affects recoil spectra much less and therefore shall be disregarded in the following.

\par The input from nuclear physics amounts to the parameters $\Delta q^{(p,n)}$ in SD scattering and $f_{T_q}^{(p,n)}$, $f_{TG}^{(p,n)}$, $q^{(p,n)}(2)$ and $\bar{q}^{(p,n)}(2)$ in the SI case. The coefficients $f_{T_q}^{(p,n)}$ and $f_{TG}^{(p,n)}$ can furthermore be expressed in terms of the $\pi-$nucleon sigma term $\Sigma_{\pi n}$ and the parameter $\sigma_0$ (see for instance \cite{EllisDD}). The fiducial values of all these quantities can be found in \cite{EllisDD,Hisano:2010ct}. Following \cite{Akrami}, the most important nuclear nuisances regard $\Delta s^{(p,n)}$, $\Sigma_{\pi n}$ and $\sigma_0$ whose uncertainties we take from \cite{EllisDD}:
\begin{eqnarray}\label{nuclear}
\Delta s^{(p)} &=& \Delta s^{(n)} = -0.09\pm 0.03 \, , \nonumber \\
\Sigma_{\pi n} &=& 64 \pm 8 \textrm{ MeV} \, , \nonumber \\
\sigma_0 &=& 36 \pm 7 \textrm{ MeV} \quad 
\end{eqnarray}
(a recent determination of $\Sigma_{\pi n}$ \cite{Alarcon:2011zs} yields $59\pm 7$ MeV, but we have checked explicitly that our results change very little using this determination instead of that in equation \eqref{nuclear}). Moreover, the parton distribution functions are particularly prone to nuisances and degrade the accuracy of SI searches. From Fig. 9 in \cite{Pumplin:2002vw} the relative uncertainty on the parton distribution function $u(x)$ ($d(x)$) is of order 10$-$25\% (10$-$50\%). Upon integration to get $u(2)$ and $d(2)$, these uncertainties are partly smoothed out. Conservatively, we consider that all the second moments in the last term of Eq.~\eqref{fp} are known to a 30\% accuracy. In the remainder of the work, all these nuclear uncertainties are folded in with the astrophysical nuisances described in the previous paragraph to quantify the total uncertainty affecting direct detection.

\subsection{Experimental data}
\par In order to derive constraints on the parameter space of our DM model, we make use of the latest results from direct detection: XENON100 \cite{Xenon100_2012} for SI scattering, XENON10 \cite{XENONSD} for SD-neutron scattering, and SIMPLE \cite{SIMPLE11} and COUPP \cite{COUPP12} for SD-proton scattering. To the best of our knowledge these are the strongest published direct detection limits for each type of couplings. It is fair to point out that the limits imposed by IceCube \cite{IceCube:2011aj} on DM-induced neutrinos from the Sun are stronger than SIMPLE and COUPP for SD-proton scattering, but they lie beyond the scope of the present paper. Recall as well that our focus here is on WIMP masses in excess of 40 GeV; data sets other than the ones mentioned above can of course be more constraining at lower masses. We briefly summarise below each experimental setup:
\begin{itemize}
\item {\bf XENON100} \cite{Xenon100_2012}. Using Xe as target material, the XENON100 experiment has set the strongest direct detection limits to date by collecting 2 WIMP-like events in the energy window $E_R=6.6-30.5$ keV for an effective exposure of 2323.7 kg.day. The corresponding background estimate is 1.0$\pm$0.2 events. Following the Feldman-Cousins procedure \cite{FeldmanCousins} with $N_{obs}=2$ observed events and $N_{bkg}=1.0$ mean expected background, we derive a 95\% CL upper limit on the WIMP signal and obtain $N_R\leq 5.72$. Notice that, besides its superb sensitivity to SI couplings due to the high number of nucleons, xenon also presents good SD-neutron sensitivity through the abundant odd isotopes $^{129}$Xe and $^{131}$Xe.

\item {\bf XENON10} \cite{XENONSD}. Early results from the XENON10 experiment corresponding to an effective exposure of 136 kg.day show $N_{obs}=10$ events in the WIMP signal region within the energy range $E_R=4.5-27$ keV. Since no background estimate is available, we stand on the conservative side and assume zero background, $N_{bkg}=0$. Then, the Feldman-Cousins procedure yields an upper limit $N_R\leq 17.82$ at 95\% CL.

\item {\bf SIMPLE} \cite{SIMPLE11}. The SIMPLE experiment consists of a threshold bubble device using C$_2$ClF$_5$ as target material. We focus here on the results of phase II which comprises stages 1 and 2, with a combined after-cuts exposure of (13.47+6.71) kg.day. For this exposure, $N_{obs}=14+1$ WIMP-like events were registered, while the neutron expected background amounted to 0.896 counts/kg/day in stage 1 and 0.253 counts/kg/day in stage 2 yielding $N_{bkg}=12.07+1.70$. Since the threshold recoil energy for bubble nucleation is $E_{thr}=8$ keV in this experimental setup, we consider a range $E_R=8-100$ keV for the derivation of our limits along with a bubble efficiency $\eta'=1-\exp\left(-\Gamma\left(E_R/E_{thr}-1\right)\right)$ where $\Gamma=4.2$. Applying the Feldman-Cousins approach with $N_{obs}=15$ and $N_{bkg}=13.77$ leads to an upper limit $N_R\leq 10.54$ at 95\% CL. This is a somewhat conservative limit because we do not address the problematics of double-scatterings, but it is sufficiently strong for our purposes.

\item {\bf COUPP} \cite{COUPP12}. Recently, the COUPP experiment -- consisting of a CF$_3$I bubble chamber -- has released results on WIMP searches for three distinct threshold energies, namely $E_{thr}=7.8,~11.0$ and 15.5 keV with total exposures $70.6,~88.5$ and 394.0 kg.day, respectively. The latter run turns out to be the most constraining for our purposes, and consequently we shall focus on such experimental setup where $N_{obs}=8$ WIMP-like events were observed. Following the analysis presented by the COUPP collaboration \cite{COUPP12}, a single-event efficiency of 79.1\% is used to correct for the effective exposure and the bubble efficiency $\eta'$ is set to unity for I recoils above threshold and to 0.49 for C and F recoils above threshold. Similarly to the case of SIMPLE, an energy window spanning $E_R=15.5-100$ keV is considered. Using the conservative hypothesis of zero background $N_{bkg}=0$, we employ the Feldman-Cousins procedure with $N_{obs}=8$ to derive $N_R\leq 15.29$ at 95\% CL.
\end{itemize}

\begin{figure*}[htp]
\centering
\includegraphics[width=0.32\textwidth]{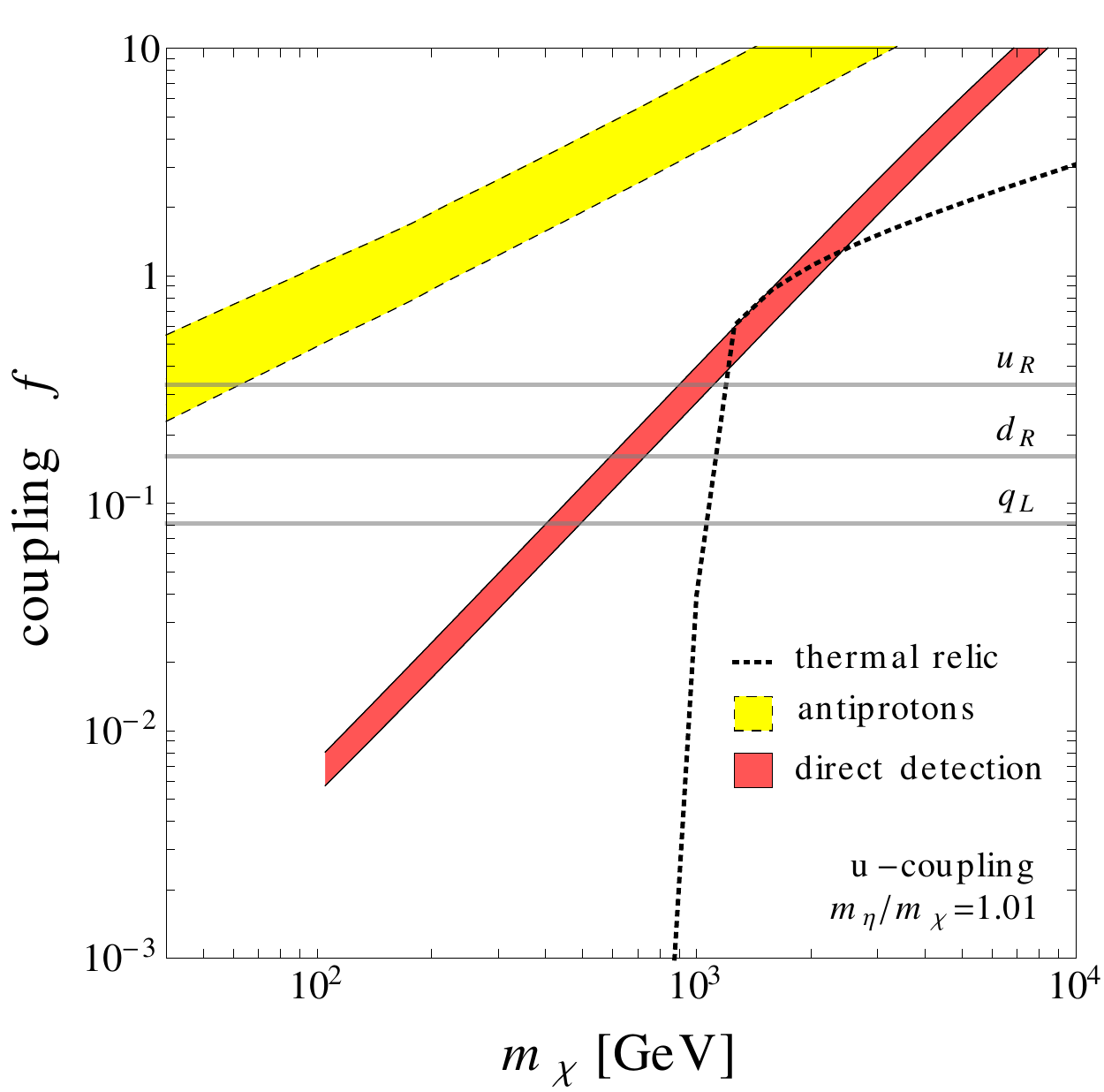}
\includegraphics[width=0.32\textwidth]{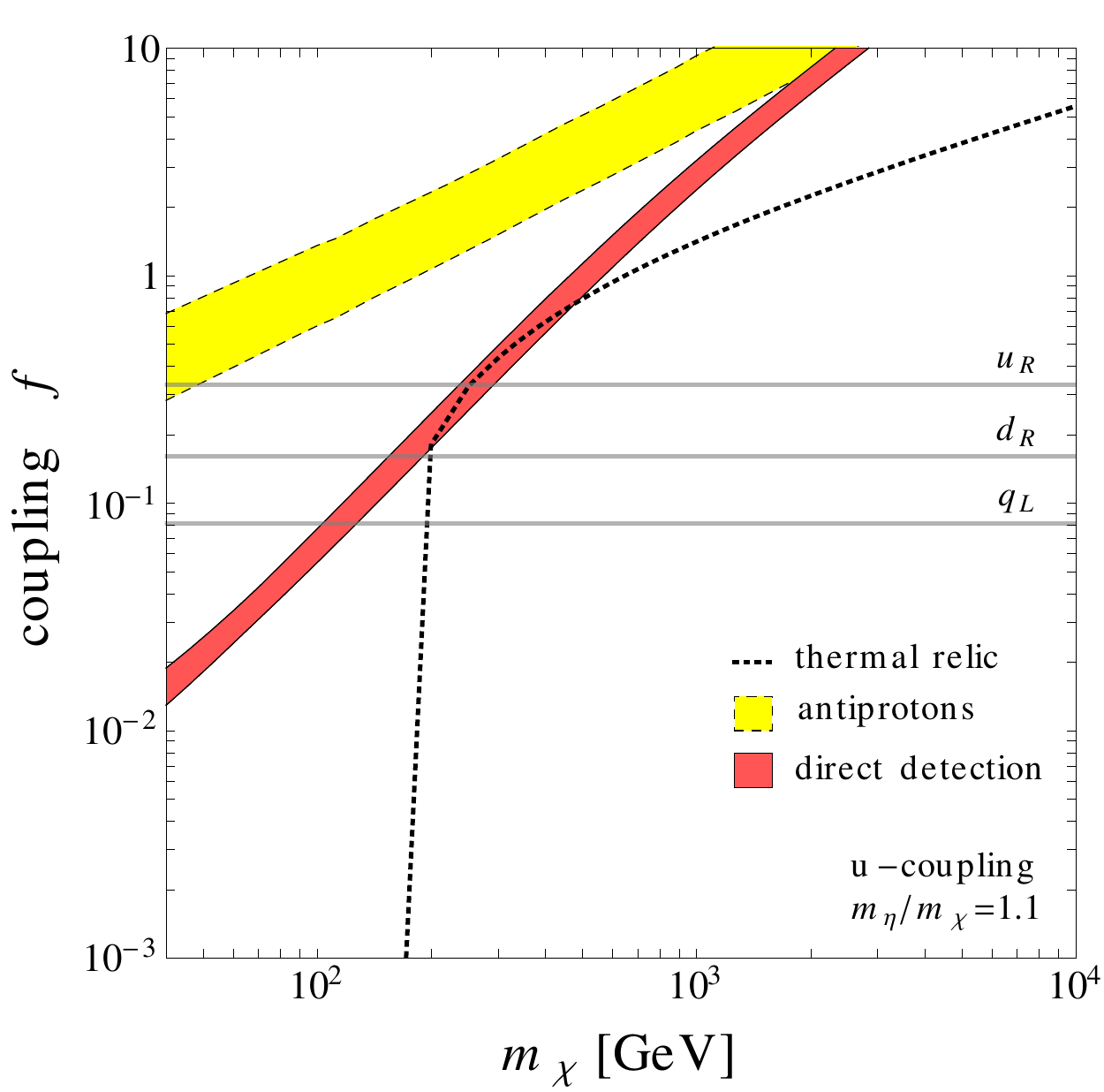}
\includegraphics[width=0.32\textwidth]{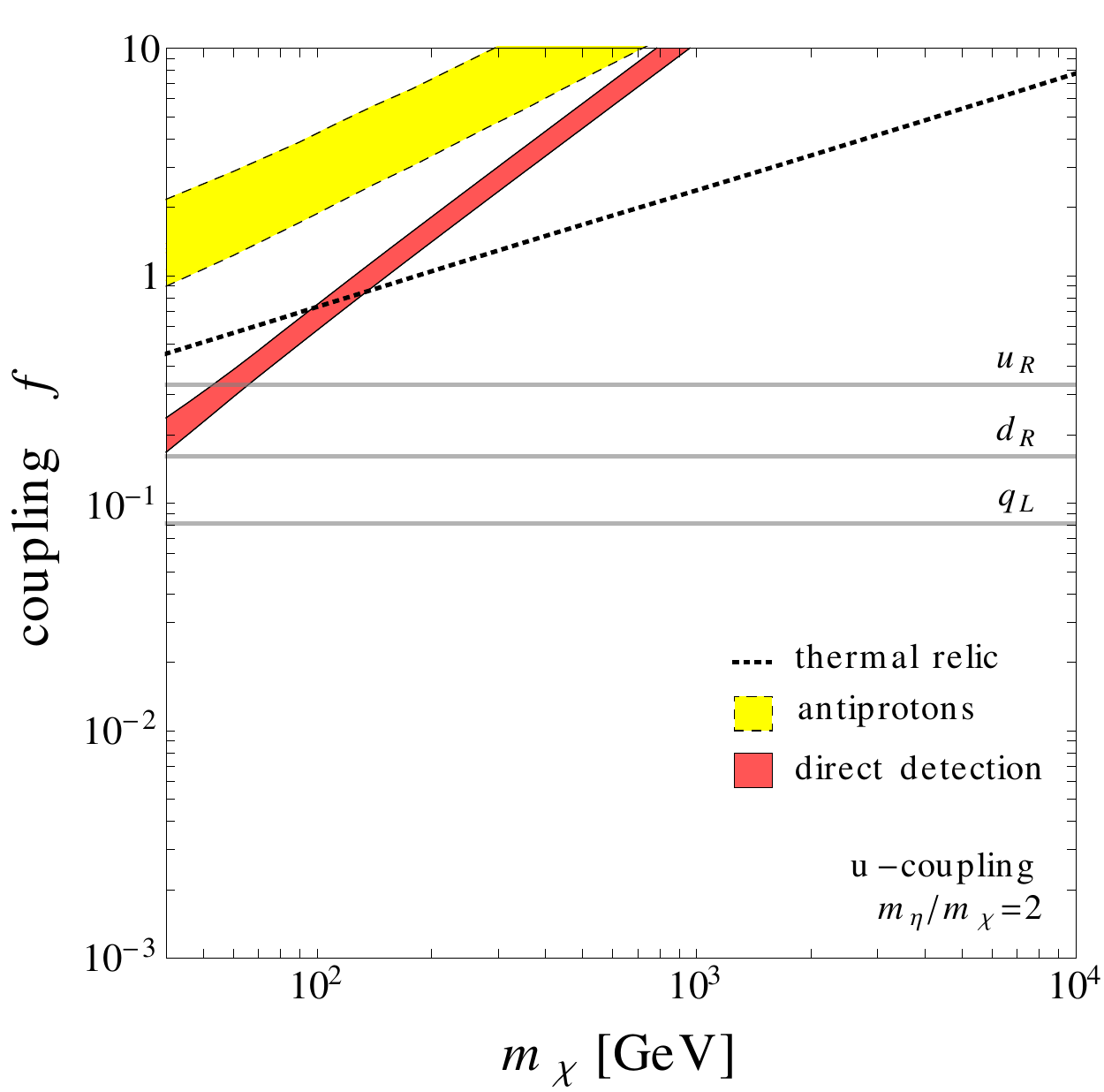}
\caption{95\% CL constraints on $f$ for the case of up-quark couplings and mass splittings $m_\eta/m_\chi=1.01,1.1,2$. The yellow band encompassed by the dashed lines represents the antiproton constraints and corresponding uncertainties, while the direct detection limits are shown by the red band and solid lines. The thick dotted lines indicate the coupling $f$ required to have a dark matter particle with the relic abundance measured by WMAP. Also shown as horizontal grey lines are the couplings of the supersymmetric bino to right-handed up-quarks, right-handed down-quarks and left-handed quarks.
}
\label{fig:coupling}
\end{figure*}

\section{Results \& discussion}\label{sec:res}

\par We present in this section the results of our numerical analysis comparing the limits on the parameter space of our toy model which stem from the non-observation of an antiproton excess in the PAMELA cosmic antiproton-to-proton fraction or a signal in direct detection experiments. We show in Fig.~\ref{fig:coupling} the limits on the coupling $f$ as a function of the dark matter mass coming from antiprotons (yellow band, dashed lines) and from direct detection (red band, solid lines), for various values of $m_{\eta}/m_{\chi}$ and assuming that the Standard Model fermion is a right-handed up-quark. The bands were constructed by marginalising over the uncertainties discussed in Sections \ref{sec:Indirect} and \ref{sec:DD} at each value of $m_\chi$. In the case of direct detection, we chose the most constraining data set at each $m_\chi$, which turned out to be always XENON100 for the dark matter models studied here. Our numerical calculation shows that there exists very stringent limits on the coupling constant from direct detection experiments when the dark matter particle is close in mass to the intermediate scalar particle. Whereas in the non-degenerate case, $m_\eta/m_\chi=2$, the direct detection upper limit on the coupling is $f\lesssim 0.8~(10)$ for $m_\chi=100~(1000)$ GeV, in the degenerate case these limits get strengthened to  $f\lesssim 0.08~(3)$ in the case $m_\eta/m_\chi=1.1$ and $f\lesssim 0.008~(0.4)$ in the case $m_\eta/m_\chi=1.01$. As a comparison, in the MSSM the Yukawa coupling of the bino to the up-quark and its corresponding squark is $f_{SUSY}\sim 0.33$, which would translate into a lower limit on the bino dark matter mass $m_\chi\gtrsim 45\textrm{ GeV},~215\textrm{ GeV}$ and 830 GeV for $m_\eta/m_\chi=2,~1.1$ and 1.01, respectively, as clear in Fig.~\ref{fig:coupling}. Notice that the quoted behaviour of the upper limits on $f$ with $m_\eta/m_\chi$ is rather strong if translated to cross-sections since $\sigma_p^{SI}\propto f^4$.

\begin{figure}[htp]
\centering
\includegraphics[width=0.49\textwidth]{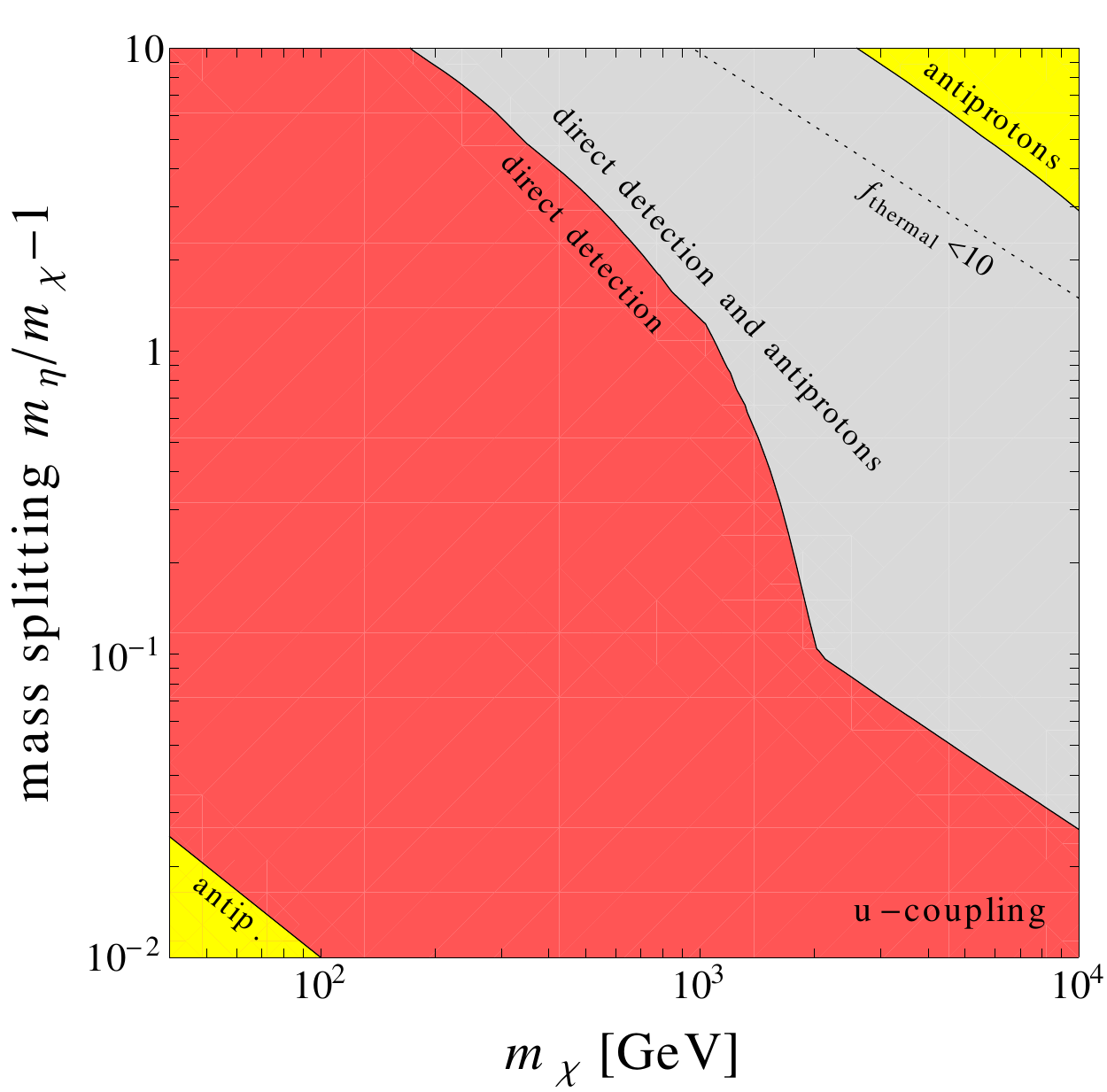}
\caption{Most dominant constraints on the mass splitting vs dark matter mass plane for the case of up-quark couplings. The red (yellow) region signals the parameter space where direct detection is more (less) constraining than antiprotons. The grey region shows the region where direct detection and antiprotons are equally important within uncertainties. In the patch above and to the right of the dotted line thermal candidates present coupling constants in excess of 10.
}
\label{fig:dominant}
\end{figure}

\begin{figure*}[htp]
\centering
\includegraphics[width=0.32\textwidth]{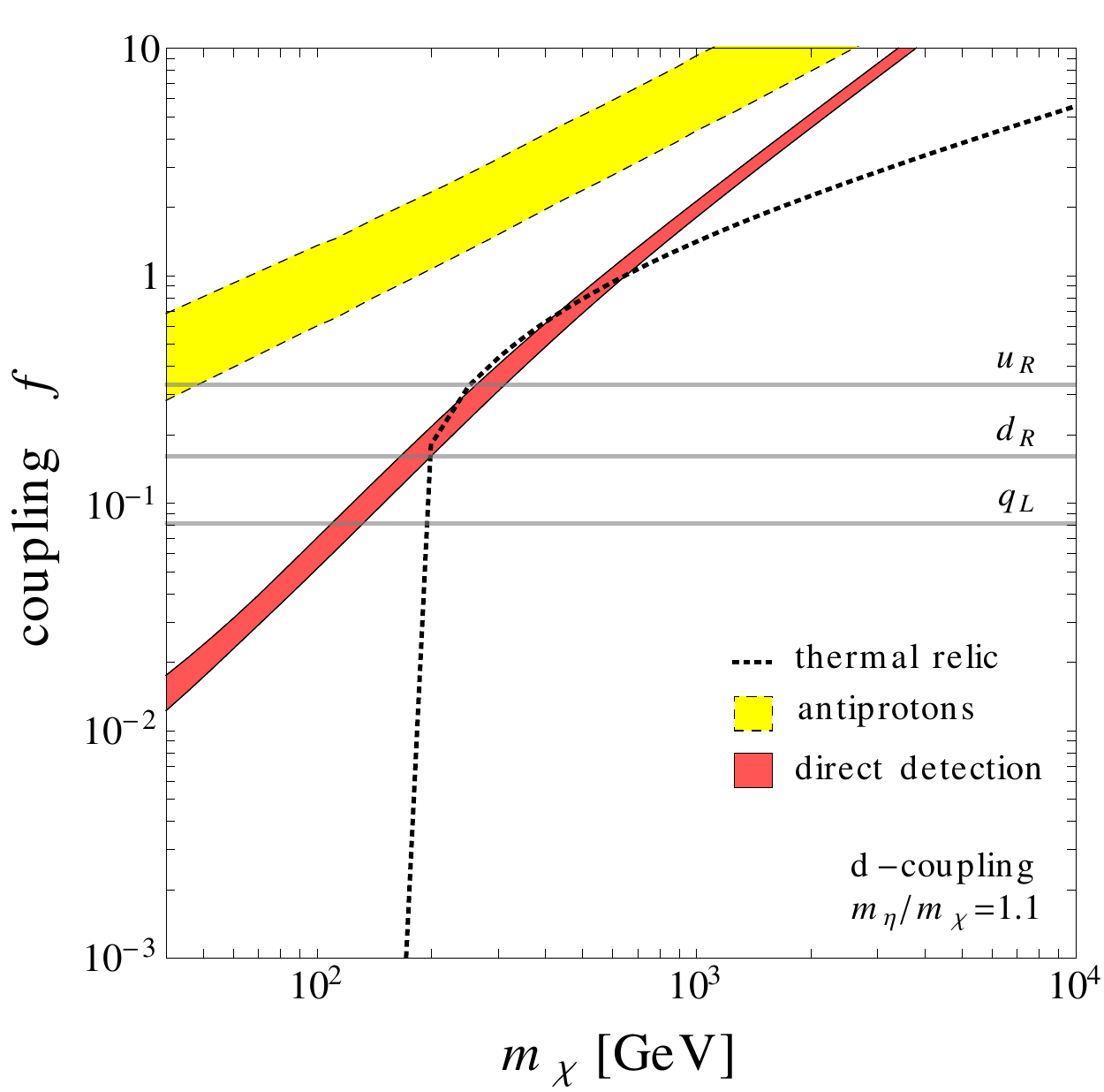}
\includegraphics[width=0.32\textwidth]{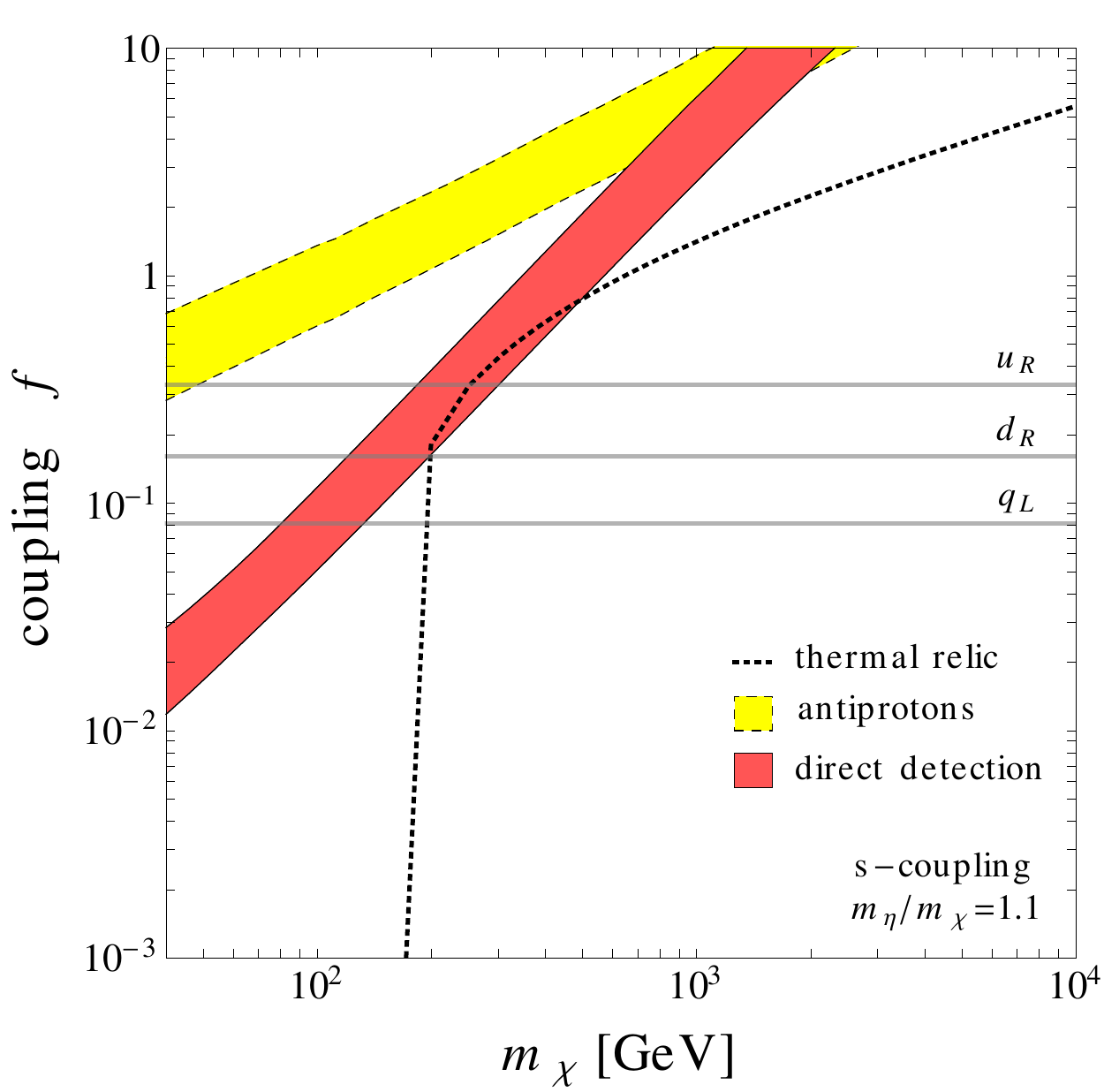}
\includegraphics[width=0.32\textwidth]{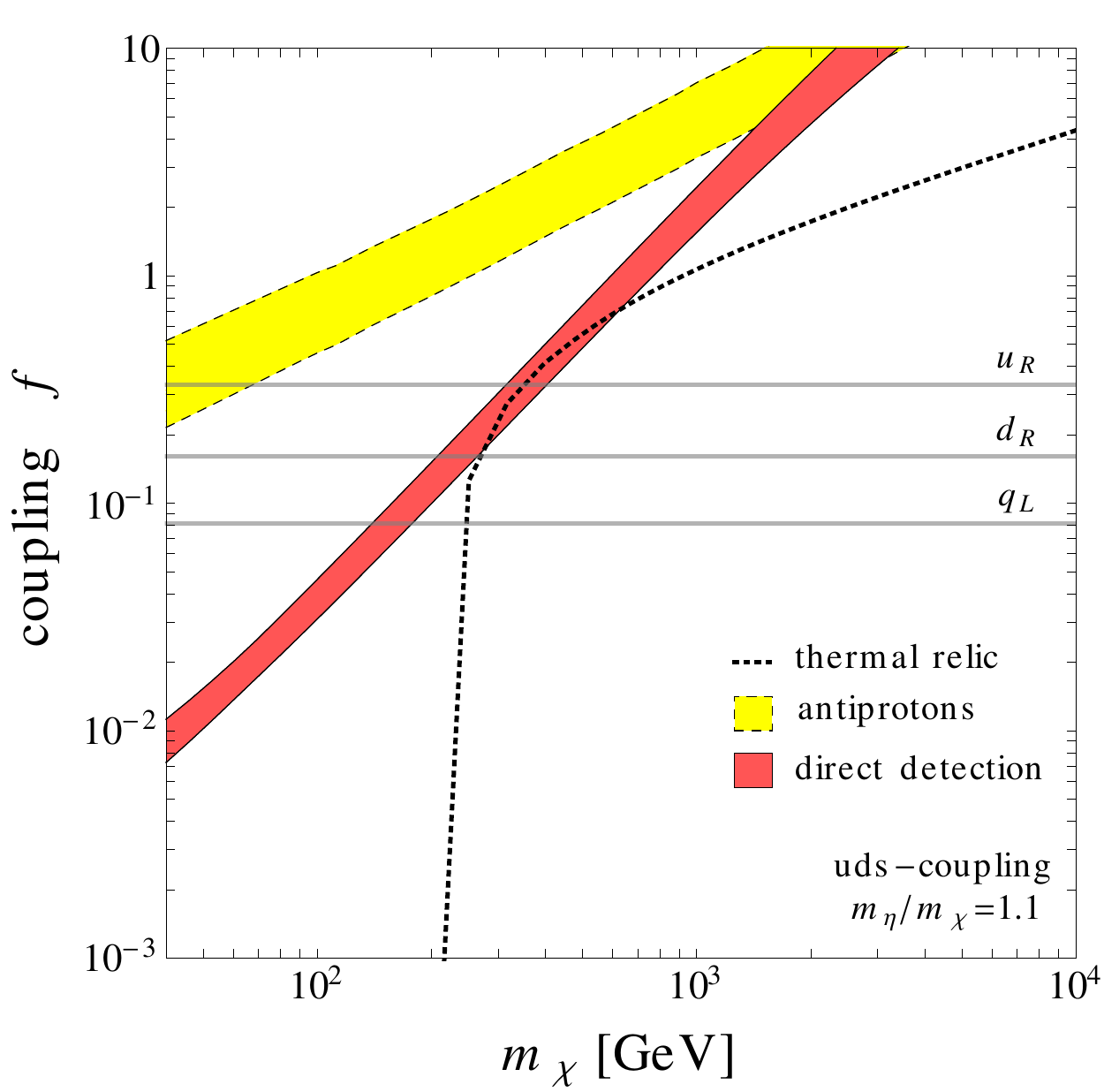}
\includegraphics[width=0.32\textwidth]{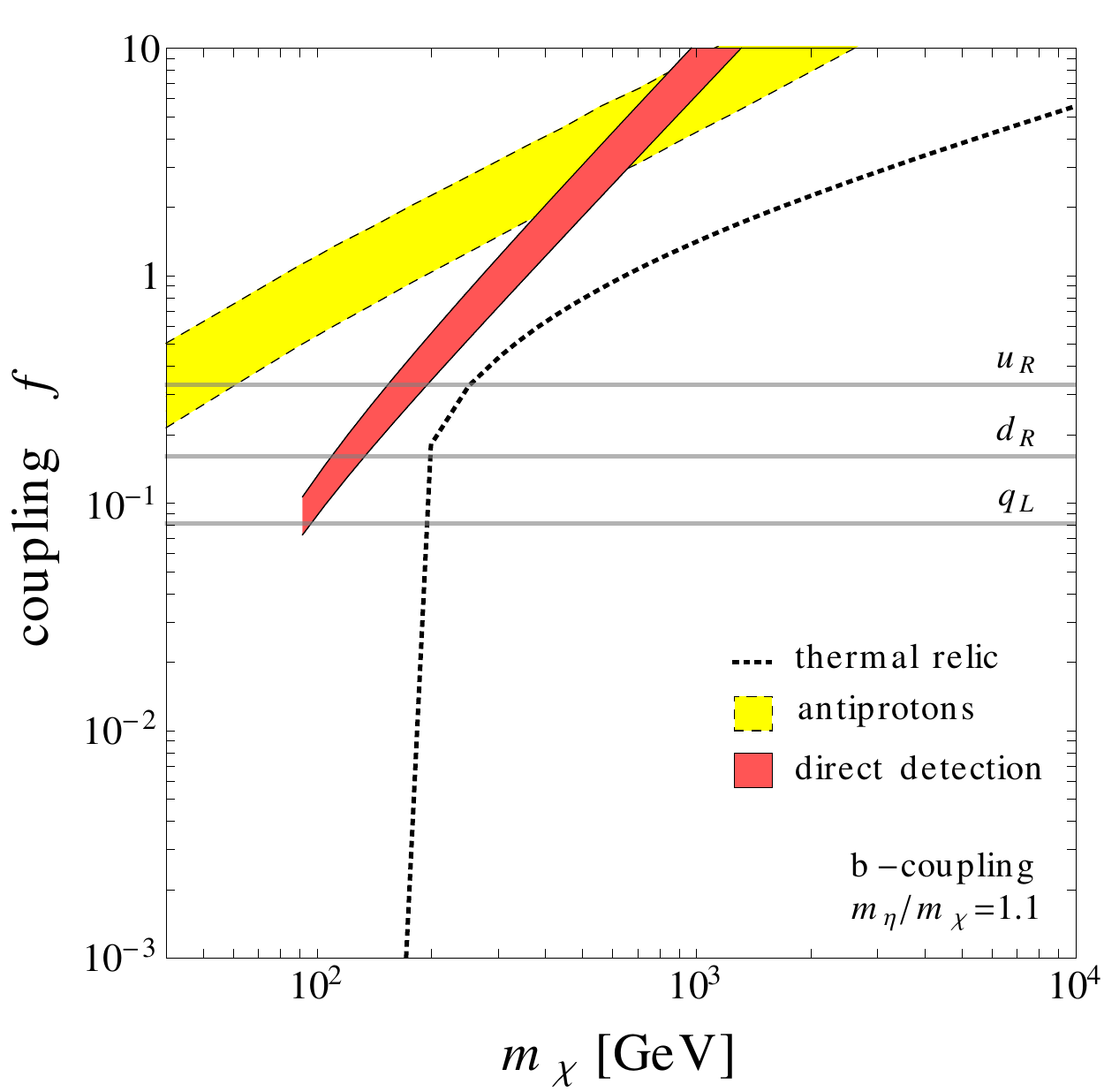}
\caption{The same as in Fig.~\ref{fig:coupling} but for a mass splitting $m_\eta/m_\chi=1.1$ and several coupling configurations.
}
\label{fig:flavour}
\end{figure*}

\begin{figure*}[htp]
\centering
\includegraphics[width=0.49\textwidth]{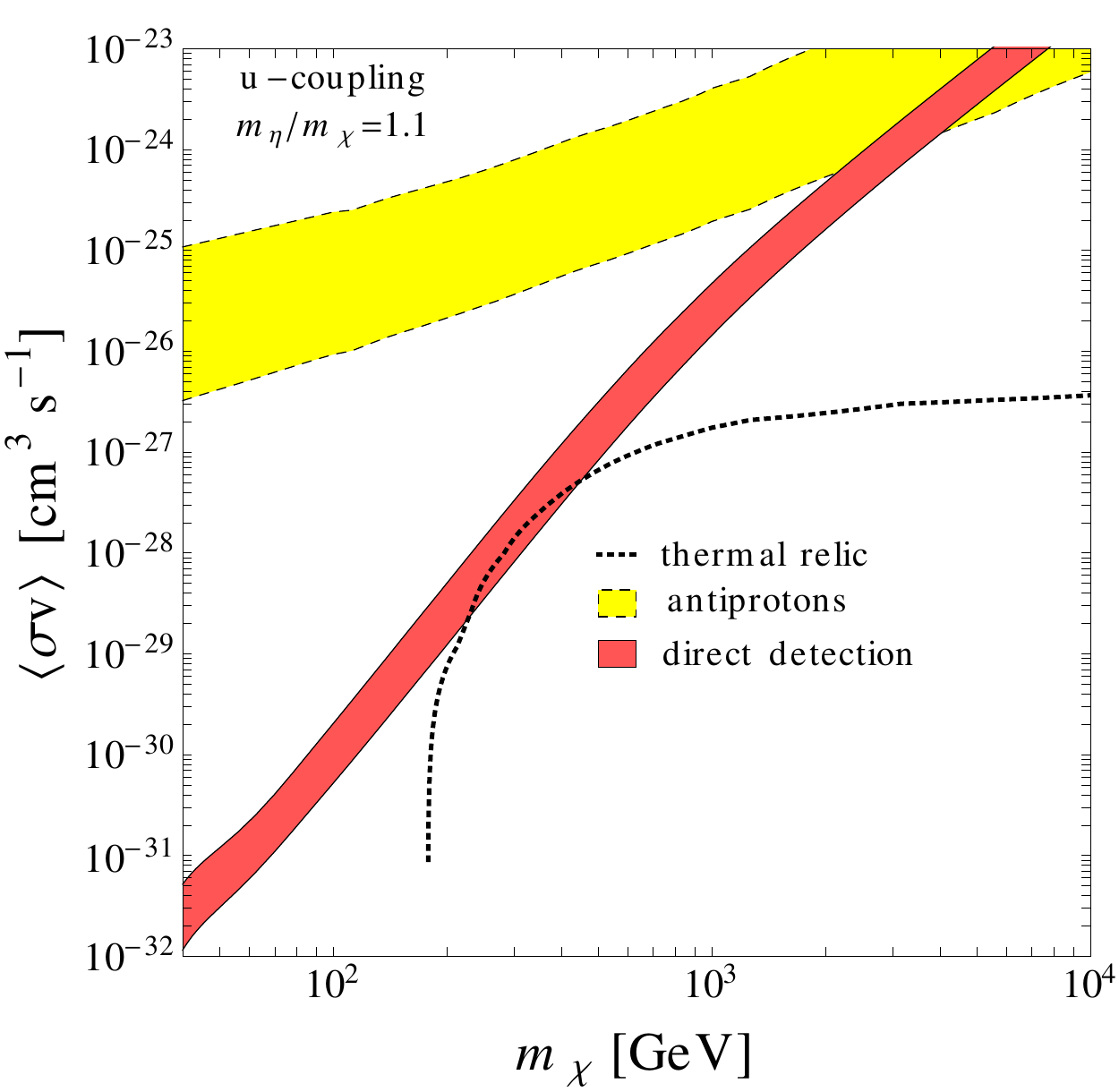}
\includegraphics[width=0.49\textwidth]{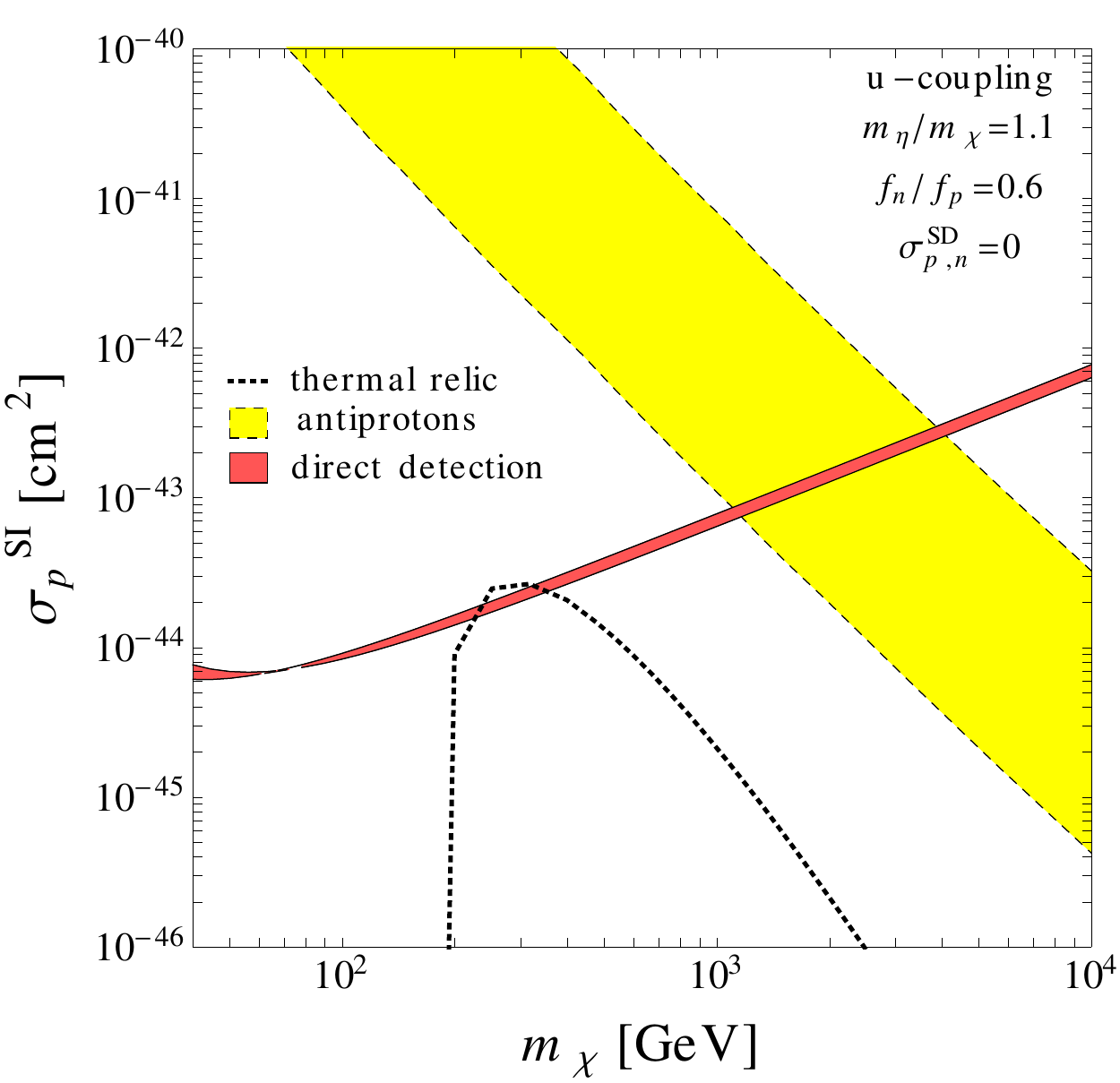}
\includegraphics[width=0.49\textwidth]{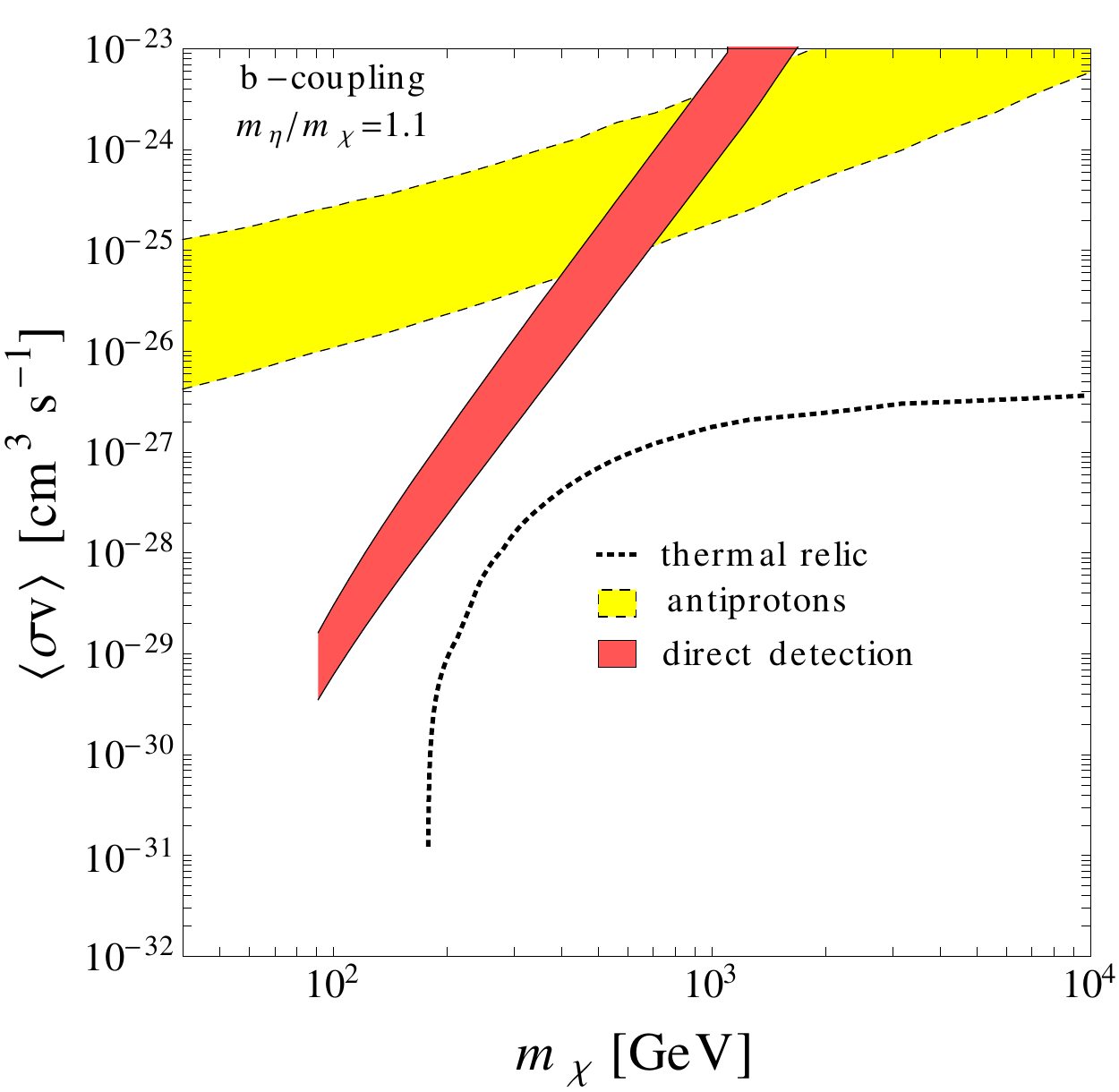}
\includegraphics[width=0.49\textwidth]{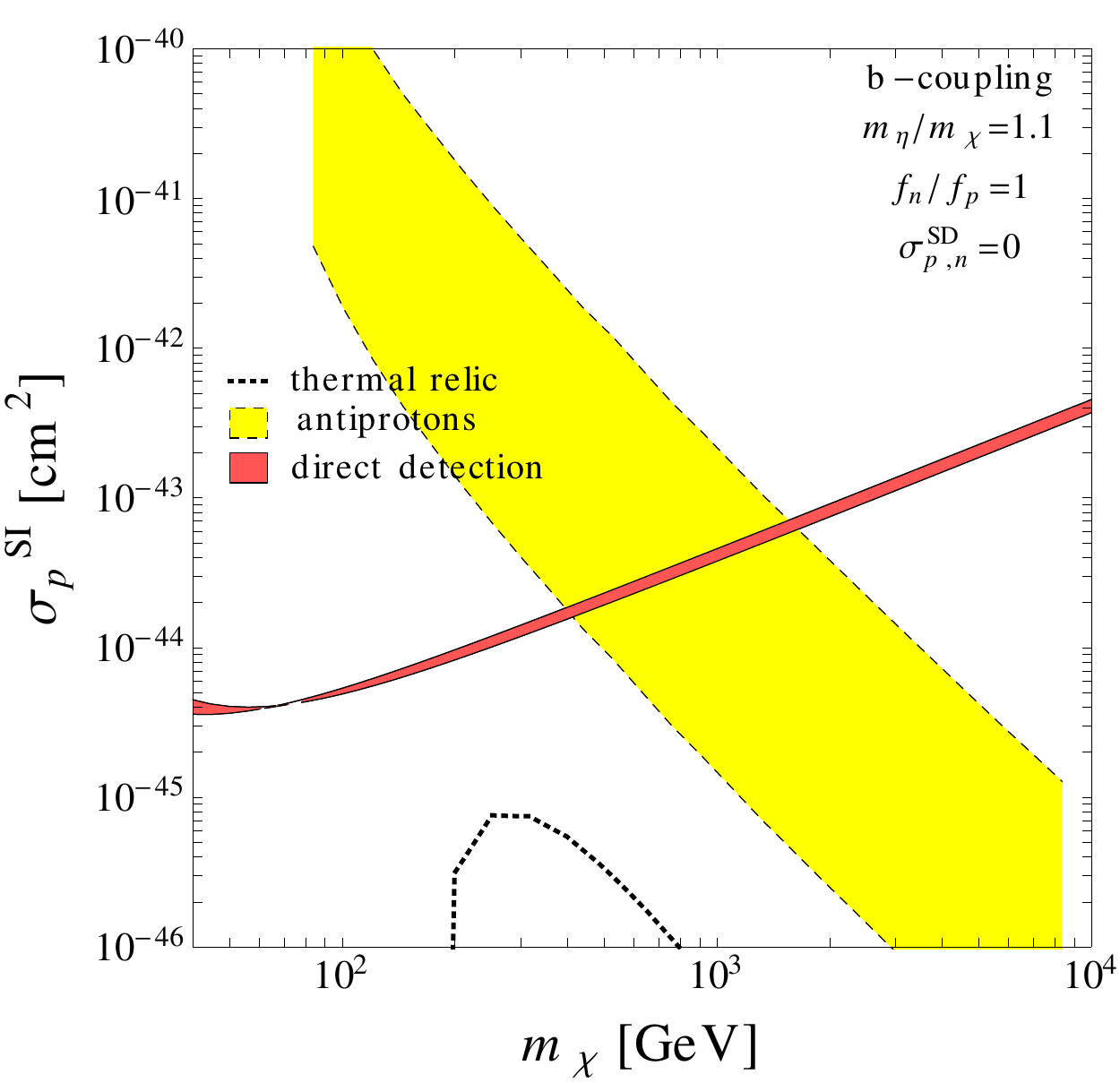}
\caption{Antiproton and direct detection 95\% CL constraints on the total annihilation cross-section (left) and on the spin-independent WIMP-proton scattering cross-section (right) for the case of up-quark (top) and bottom-quark (bottom) couplings and a mass splitting $m_\eta/m_\chi=1.1$. The colour and line code is the same as in Fig.~\ref{fig:coupling}.
}
\label{fig:DDantip}
\end{figure*}

\begin{figure}[htp]
\centering
\includegraphics[width=0.49\textwidth]{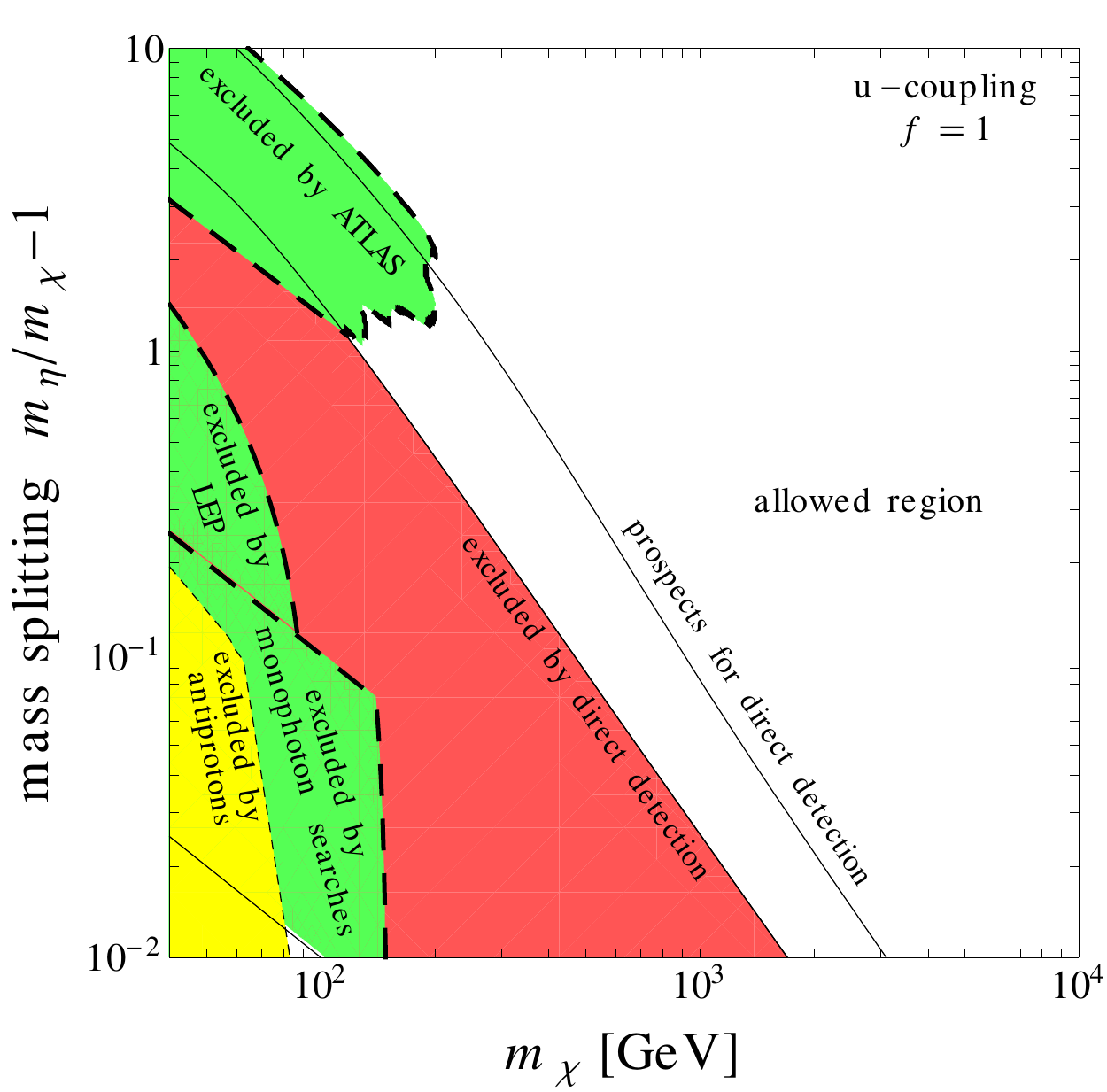}
\caption{95\% CL exclusion regions on the mass splitting vs dark matter mass plane for the case of up-quark couplings and fixing $f=1$. The solid red, thin dashed yellow and thick dashed green contours correspond to the regions of the parameter space excluded by direct detection, antiprotons and collider searches, respectively. The upper solid contour shows the reach of sensitivity of XENON1T.
}
\label{fig:compl}
\end{figure}

\par The limits on the parameter space of the model from antiprotons are generically weaker than those from direct dark matter searches. This can be explicitly seen in Fig.~\ref{fig:dominant}, where we plot the relative importance of each constraint in the mass splitting vs dark matter mass plane. The red region delimits the parameter space for which direct detection limits take the lead over antiprotons even when uncertainties are taken into account -- i.e.~when the most conservative direct detection bound is stronger than the most aggressive antiproton constraint. The yellow regions refer to the opposite case, while the grey zone labelled ``direct detection and antiprotons'' marks the region where antiprotons and direct detection are equally important within uncertainties. Only at large values of the dark matter mass the limits from antiprotons become more important, due to the different scaling of the limits with the mass. However, with the current data, this occurs for values of the coupling constant where our perturbative calculation is no longer valid -- see dotted line in Fig.~\ref{fig:dominant}. Furthermore, as pointed out in Section \ref{sec:DD}, we note that when the dark matter mass is close to the mass of $\eta$, the perturbative calculation of the direct detection rate breaks down, which happens at $m_\chi\lesssim 100\GeV$ for $m_\eta/m_\chi=1.01$, hence in this regime only the antiproton limits are reliable.

\par We also show in the plots of Fig.~\ref{fig:coupling}, as a dotted line, the coupling $f$ required to give the dark matter thermal relic density as measured by WMAP. As discussed in Section \ref{sec:model}, this requirement sets a lower limit on the dark matter mass, which increases as $m_\eta/m_\chi$ approaches 1, namely $m_\chi\gtrsim {\cal O}[200~(1000)\GeV]$ for $m_\eta/m_\chi=1.1~(1.01)$. Above the dotted line in Fig.~\ref{fig:coupling}, additional production mechanisms should be present, such as non-thermal production, in order to reproduce the observed dark matter relic density. As can be seen from Fig.~\ref{fig:coupling}, in the degenerate case direct detection experiments start to probe the region of the parameter space favoured by thermal production and, in the non-degenerate case, even exclude the region $m_\chi\lesssim 100\GeV$.

\par To examine the sensitivity of these conclusions to the flavour of the quark to which dark matter couples, we have repeated the same analysis for couplings to the down-quark, the strange-quark, the bottom-quark and a democratic coupling to the up-, down- and strange-quarks; the results are shown in Fig.~\ref{fig:flavour} for the case $m_\eta/m_\chi=1.1$. The conclusions for the three light quark flavours  are very similar, the main difference being the width of the uncertainty band in the direct detection limits, which is larger in the case of the strange-quark. Furthermore, the antiproton and direct detection limits for the scenario where the dark matter particle couples democratically to the three light quark flavours are similar to those for the coupling to the individual flavours. Lastly, we have also calculated the limits on the coupling in a scenario where the dark matter particle only couples to the bottom-quark. While the antiproton limits are only slightly stronger than those for the couplings to the light quark flavours, the direct detection limits are significantly weaker. Still, the limits on this scenario from direct detection are stronger than from antiprotons. This is a particularly strong conclusion given the numerous works in the literature studying the complementarity between indirect and direct detection assuming WIMP annihilations into pairs of bottom-quarks.

\par The previous limits on the coupling $f$ can also be translated into upper limits on $\sigma_p^{SI}$ and $\langle \sigma v \rangle$. The former translation suffers, of course, from the nuclear uncertainties discussed in Section \ref{sec:DD}. The limits from direct detection and antiprotons are presented in Fig.~\ref{fig:DDantip} as a red and a yellow band respectively, for the case $m_\eta/m_\chi=1.1$ and couplings to up- and bottom-quarks. Note that in the right plots of Fig.~\ref{fig:DDantip} we neglect SD scattering in order to show the familiar SI parameter space -- strictly speaking, for up-quark couplings and $m_\eta/m_\chi=1.1$, in an experiment such as XENON100 this approximation is valid up to masses of a few TeV, above which SD scattering becomes important. As mentioned above, for low dark matter masses direct detection experiments provide the most stringent constraints on the model, while for large masses antiproton searches become dominant. However, we stress that the region at large masses where the antiproton limits are stronger than the direct detection limits corresponds to couplings larger than 10, hence the perturbative calculation is no longer reliable. Moreover, it should be borne in mind that the bands shown in Fig.~\ref{fig:DDantip} are sensitive to astrophysical quantities: the antiproton limits can become stronger in the presence of ``boost factors'' and the direct detection limits can become stronger or weaker depending on the concrete value of the dark matter density at the solar system. We finally notice that the gamma-ray and antiproton limits on annihilations into quarks are of similar strength \cite{Bringmann:2012vr}, and therefore any interpretation of the gamma-ray feature around 130 GeV as internal bremsstrahlung $\chi\chi \to q\bar{q}\gamma$ is in tension with direct detection constraints.

\par We now turn to the complementarity of antiproton and direct detection constraints in light of collider searches. For concreteness, we shall focus on the case of a WIMP coupling to up-quarks only. Fig.~\ref{fig:compl} shows the regions in the mass splitting vs dark matter mass plane excluded by direct detection (red, solid) and antiprotons (yellow, thin dashed) assuming a coupling $f=1$ and taking the most conservative limits -- i.e.~the upper lines in the bands of Fig.~\ref{fig:coupling}. At $m_\chi=300~(1000)$ GeV direct detection excludes mass splittings $m_\eta/m_\chi-1\lesssim 19~(2)\%$ for $f=1$ and $6~(<1)\%$ for $f=f_{SUSY}=0.33$. Notice that the direct detection contour towards the bottom left corner of Fig.~\ref{fig:compl} signals the limit of validity of our calculations ($m_\eta-m_\chi>1$ GeV); the region below this line, although not strictly excluded by direct detection in our analysis, should be regarded as very contrived. Overplotted to the antiprotons and direct detection limits in Fig.~\ref{fig:compl} is the region excluded by ATLAS (green, thick dashed) in the search for jets plus missing transverse momentum \cite{ATLAS-CONF-2011-155}. Previous limits from LEP and Tevatron are relevant for dark matter masses $m_\chi\lesssim 100$\,GeV and mass splittings smaller than the region probed by ATLAS. For scalar quarks of the first generation, L3 \cite{Achard:2003ge} has put a limit $m_\eta\geq 97$\,GeV for $m_\eta-m_\chi\geq 10$\,GeV, which is shown in Fig.~\ref{fig:compl}\footnote{Note that this bound is obtained assuming degenerate squarks of the first two generations. Relaxing this assumption would slightly reduce the mass lower limit by a few GeV.}. For the case of a scalar bottom quark (not shown here), D0 and CDF limits \cite{Abazov:2010wq,Aaltonen:2010dy} require $m_\eta\gtrsim 200$\,GeV for $m_\eta/m_\chi\gtrsim 1.5$ and $m_\chi\lesssim 100$\,GeV. LEP searches instead exclude scalar bottom quarks with masses $m_\eta\lesssim 95$\,GeV for $m_\eta-m_\chi=20$\,GeV \cite{sboldLEP}. Additionally, a recent re-analysis \cite{Belanger:2012mk} of monophoton plus missing transverse energy LHC data derives a lower limit $m_\eta\gtrsim 150$ GeV for up-type quarks provided $m_\eta-m_\chi<10$ GeV. The latter constraints, however, only probe a small patch in the bottom left corner of Fig.~\ref{fig:compl} which is largely excluded by direct detection if $f=1$.  Fig.~\ref{fig:compl} clearly shows that collider searches are effectively orthogonal to antiproton and direct detection limits. In fact, although the allowed region is large for high masses, in the low-mass regime only a small region still survives. This underlines the importance of pursuing complementary searches to close in on mass-degenerate dark matter models.

\par It is therefore interesting to investigate the prospects for antiproton, direct and collider searches. We start with antiprotons. The AMS-02 detector \cite{ams02site} -- installed on the International Space Station in 2011 -- is presently collecting data and will likely extend the measurement of the antiproton flux to higher energies than PAMELA. However, it will probably take some time to reach the accuracy of PAMELA antiproton-to-proton data, which is the data we employ in the present analysis. Therefore we assume that antiproton constraints on dark matter models will not improve significantly in the short term. Direct detection, instead, presents promising prospects with several ton-scale experiments \cite{darwin,eureca} already on the way. We focus on a relatively short time scale and take XENON1T as an example: it is expected \cite{Baudis:2012bc} that XENON1T will improve a factor $\sim 60$ in SI cross-section sensitivity with respect to the latest XENON100 results \cite{Xenon100_2012}. As shown in Fig.~\ref{fig:compl}, such prospects will effectively shift the direct detection constraints (which are dominated by XENON100) upwards and, assuming a coupling $f = 1$, these will cover the reach of current LHC searches. It is clear that a big chunk of the parameter space of mass-degenerate models will be excluded over the next few years if no signal is detected. In that case we expect stringent lower limits on the mass splitting of order 114\% (10\%) at $m_\chi=300~(1000)$ GeV for $f=1$.

\section{Conclusion}\label{sec:conc}
\par We have shown in this work that the latest data on cosmic antiprotons and direct detection place useful constraints on the phenomenology of mass-degenerate dark matter scenarios. In particular, we have considered a minimal framework featuring a Majorana fermion as dark matter that couples to light quarks via a scalar close in mass, encompassing e.g.~a simplified model with bino-like neutralino and squark as lightest and next-to-lightest supersymmetric particles. This setup allows for a direct comparison of scattering rates on nuclei with dark matter annihilation rates in our Galaxy, the dominant channel being $\chi\chi\to q\bar q g$ for quarks of the first and second generation. The derived constraints on coupling to quarks suffer from sizeable astrophysical and nuclear uncertainties, but it is nevertheless clear that antiprotons lag significantly behind direct detection, a fact that can be attributed mainly to the extreme sensitivity of underground searches to mass degeneracy. Fine degeneracies are conclusively discarded by current direct detection data. This is precisely the range that escapes detection at collider searches. Accordingly, we find that the interplay between antiprotons, direct and collider searches will be of crucial importance in closing in on simple mass-degenerate dark matter models over the coming years. Further work is needed to study the implications of this complementarity in the framework of more complicated particle physics realisations.

\vspace{0.5cm}
{\it Acknowledgements:} The authors thank Junji Hisano for helpful discussions and Jose Manuel Alarc\'on, Germano Nardini and Pat Scott for useful comments. This work has been partially supported by the DFG cluster of excellence ``Origin and Structure of the Universe'' and by the DFG Collaborative Research Center 676 ``Particles, Strings and the Early Universe''. S.V.~also acknowledges support from the DFG Graduiertenkolleg ``Particle Physics at the Energy Frontier of New Phenomena''.

%%%%%%%%%%%%%%%%%

\bibliographystyle{apsrev}%4-1.bst}
\bibliography{DDpbar}

\end{document}